\documentclass[a4paper]{jpconf}
\usepackage{graphicx}
\usepackage{subfig}
\usepackage{amsmath}	
\usepackage{amssymb}	
\begin{document}
\title{Test particles in relativistic resistive magnetohydrodynamics}

\author{Bart Ripperda$^{1}$, Oliver Porth$^{2}$ and Rony Keppens$^{1}$}

\address{$^{1}$ Centre for mathematical Plasma Astrophysics, Department of Mathematics, KU Leuven, Celestijnenlaan 200B, B-3001 Leuven, Belgium}
\address{$^{2}$ Institut fur Theoretische Physik, Max-von-Laue-Str. 1, D-60438 Frankfurt, Germany}

\ead{bart.ripperda@kuleuven.be}

\begin{abstract}

The Black Hole Accretion Code ({\tt BHAC}) has recently been extended with the ability to evolve charged test particles according to the Lorentz force within resistive relativistic magnetohydrodynamics simulations. We apply this method to evolve particles in a reconnecting current sheet that forms due to the coalescence of two magnetic flux tubes in 2D Minkowski spacetime. This is the first analysis of charged test particle evolution in resistive relativistic magnetohydrodynamics simulations.
The energy distributions of an ensemble of 100.000 electrons are analyzed, as well as the acceleration of particles in the plasmoids that form in the reconnection layer. The effect of the Lundquist number, magnetization, and plasma-$\beta$ on the particle energy distribution is explored for a range of astrophysically relevant parameters. We find that electrons accelerate to non-thermal energies in the thin current sheets in all cases. We find two separate acceleration regimes: An exponential increase of the Lorentz factor during the island coalescence where the acceleration depends linearly on the resistivity and a nonlinear phase with high variability. These results are relevant for determining energy distributions and acceleration sites obtaining radiation maps in large-scale magnetohydrodynamics simulations of black hole accretion disks and jets.

\end{abstract}

%
%
%

\section{Introduction}

Relativistic magnetic reconnection is considered to be the main driving mechanism behind the characteristic energetic flaring activity from accreting compact objects. Instabilities occurring in the accretion disk plasma can produce current sheets where reconnection is triggered. Such reconnection layers are susceptible to the plasmoid instability which breaks the current sheet into a chain of magnetic islands or plasmoids. X-ray and radio emission from supermassive black holes in active galactic nuclei (AGN) and X-ray binaries are attributed to electrons that have gained their energy in reconnection zones in the coronae above the accretion disk. The particles are ejected as energetic plasmoids that are associated with the flaring activity (see e.g. \cite{kagan2014} for a review). 

Reconnection regions are potentially located nearby the black hole event horizon, such that the plasma flow has to be modelled within the framework of relativistic magnetohydrodynamics. Ideal general relativistic magnetohydrodynamics (GRMHD) simulations are often used to study the accretion disk around compact objects. Such ideal models however, assume that the plasma is infinitely conductive and that resistive dissipation plays no role in the dynamics. For global accretion disk and jet dynamics the Lundquist number $S = v_A L / \eta \approx c L / \eta$ is typically extremely high with the Alfv\'{e}n speed $v_A$ approximately equal to the speed of light $c$, $L$ a typical large astrophysical length scale and $\eta$, a small but finite resistivity. In such cases, the ideal MHD approximation $S \rightarrow \infty$ and $\eta \rightarrow 0$ seems adequate. In reconnection regions however, this approximation is inappropriate and the underlying assumption of perfect conductivity is violated such that the Lundquist number is high, but not infinite. For reconnection occurring in ideal GRMHD simulations of accretion disks resistivity is provided by uncontrollable numerical effects due to a finite resolution (e.g. \cite{ball2018}; \cite{kadowaki2018}). A reconnecting plasma layer has a rapidly varying magnetic field $\mathbf{B}$ resulting in a high current density $\mathbf{J}$ that can lead to strong Ohmic dissipation $\sim \eta J^2$ and the formation of a resistive electric field $\mathbf{E} \sim \eta \mathbf{J}$. These time-dependent effects are not captured in ideal MHD descriptions since the electric field is considered a derived variable determined from $\mathbf{E} = - \mathbf{v} \times \mathbf{B}$. In general relativistic resistive magnetohydrodynamics (GRRMHD) instead, the electric field is evolved according to Amp\`{e}re's law as an independent dynamic variable. Through Ohm's law the effect of a physical and controllable resistivity is taken into account allowing for reconnection and plasmoid formation occuring naturally.

To fully capture all aspects of reconnection resistive MHD is not sufficient either. Reconnection has been shown to be an efficient source of non-thermal particle acceleration explaining the energy emission through episodic flaring from jets and accretion disks (\cite{giannios2013}; \cite{sironi2014}; \cite{melzani2014}; \cite{li2015}; \cite{sironi2016}; \cite{nalewajko2016}; \cite{petropoulou2016}; \cite{ball2016}; \cite{werner2017}; \cite{rowan2017}; \cite{werner2018}). MHD describes the fluid properties of the plasma in thermal equilibrium and can by definition not capture the physics of non-thermal particles. The nonlinear interaction between the electromagnetic fields and the particles in the plasma is consistently solved in Particle-in-Cell (PiC) simulations. In such descriptions the plasma is considered to be collisionless and kinetic effects act as an effective resistivity, allowing for reconnection to occur. First efforts are currently being made to numerically simulate general relativistic effects with PiC simulations (\cite{watson2010}; \cite{philippov2015}; \cite{levinson2018}; \cite{bacchini2018iau}; \cite{bacchini2018b}; \cite{parfrey2018}). A fully kinetic description is however prohibitive in large-scale astrophysical simulations of accretion disks and jets. Recent advances to resolve non-thermal physics in accretion disks rely on the combination of large-scale GRMHD simulations, combined with a form of sub-grid particle information. \cite{ressler2015}, \cite{ressler2017}, \cite{sadowski2017} and \cite{ryan2017} evolve a thermal electron population alongside the other MHD variables accounting for particle heating. \cite{chael2017} employ a scheme to co-evolve a population of non-thermal electrons to analyze radiation in GRMHD simulations. None of the above works however reproduces the characteristic X-ray flares with hard energy spectra that are typically observed on a daily basis from Sagittarius A* (Sgr A*), the black hole in the Galactic Centre (see e.g. \cite{baganoff2001}; \cite{genzel2003}; \cite{eckart2006}; \cite{neilsen2013}; \cite{brinkerink2015}). \cite{ball2016} and \cite{davelaar2018} assume a combination of thermal and non-thermal electron distributions, to analyze radiation signatures from GRMHD simulations in postprocessing. Their work indicates that non-thermal electron injection is necessary for modeling the energy spectra and highly variable flaring from Sgr A*. To consistently model non-thermal emission, particle dynamics has to be included in the GRMHD evolution. \cite{ball2018}, \cite{rowan2017} and \cite{chael2018} have recently overcome this issue by using a special relativistic PiC method to provide non-thermal particle distributions in locally flat slabs of large-scale GRMHD simulations. In these ideal GRMHD simulations, reconnection however occurs due to numerical resistivity instead of a physically motivated resistivity.

Relativistic resistive MHD can describe both the large-scale accretion flow, as well as the reconnection that is induced by instabilities occurring in the accretion disk. However, resistive MHD cannot self-consistently capture particle acceleration either.
We combine relativistic resistive MHD simulations of a forming current sheet, with an ensemble of electrons, following the electromagnetic fields from the MHD, according to the relativistic Lorentz force. The test particles evolve simultaneously alongside the MHD but do not exert any feedback on the electromagnetic fields. Resistive electric fields, arising from magnetic reconnection, are efficient particle accelerators (\cite{Rosdahl}; \cite{Gordovskyy}; \cite{Pinto}; \cite{Ripperda}; \cite{Ripperda2}; \cite{akramov2017}), such that non-thermal particle distributions can naturally form in resistive MHD. The test particle assumption made here is valid for a plasma where the non-thermal particles do not dominate the plasma energetics compared to the thermal ensemble. In this way, global simulations on astrophysical scales can be carried out without the computational limitations PiC approaches usually suffer from, and particle dynamics can be analyzed within the computationally cheaper test particle approach. However, the assumption that non-thermal particles are not dynamically important becomes invalid in reconnection zones where a substantial fraction of particles typically reach high Lorentz factors. Considering kinetic feedback of these particles to the electromagnetic fields requires a PiC approach. Such a method is however currently impractical due to the large scale-separation between the gyro-motion and the global accretion flow and can therefore only be applied by assuming unrealistic magnetic field strengths for accretion disk scenarios.

We assume a scenario where a turbulent accretion disk has formed around a black hole. In the corona above the accretion disk, a force-free magnetic field is represented by an ensemble of flux loops tied to the disk (see e.g. \cite{uzdensky2008}; \cite{goodman2008}; \cite{li2017}). We focus on two such flux tubes, merging and reconnecting with each other. Since the reconnection dynamics occur in a small region of the corona, we assume a flat slab of spacetime, such that we can ignore the effects of gravity and curved spacetime. The formation of a reconnection sheet in between two merging flux tubes has been investigated by \cite{SironiPorth} with special relativistic PiC simulations and by \cite{ripperda2018coalescence} with special relativistic resistive magnetohydrodynamics (SRRMHD) simulations. Here, we show a proof-of-principle combination of relativistic resistive MHD and particle methods. We employ the GRRMHD code {\tt BHAC} (\cite{porth2017}; \cite{olivares2018}) to study the global scale reconnection dynamics and plasmoid properties of a forming current sheet in between two coalescing Lundquist tubes in 2D Minkowski spacetime. By following an ensemble of 100.000 test particles in MHD fields, we analyze the influence of the Lundquist number, plasma-$\beta$ and magnetization $\sigma$ on acceleration sites and particle energy distributions. 

\section{Numerical setup}
Black hole and neutron star magnetospheres and their outflows are considered to be filled with conducting electron-positron plasmas (\cite{Blandford1977}; \cite{Komissarov3}; \cite{arons2012}). Therefore we evolve $N_{tot} = 100.000$ electrons with charge $q$ and mass $m$ in the high-resolution SRRMHD simulations of 2D merging Lundquist tubes of \cite{ripperda2018coalescence}. Due to the test particle approximation, positrons can be neglected as their dynamics are equivalent to electron dynamics, with the only difference being their opposite charge. We use a special relativistic Boris scheme to evolve charged test particles evolving in the electromagnetic fields obtained from MHD, according to the Lorentz force \cite{ripperda2017}. A method to moderate spurious acceleration of test particles, due to a lack of feedback on the electromagnetic fields, a too large resistivity, and periodic boundary conditions, was proposed in \cite{Ripperda} and \cite{Ripperda2}. For every particle that leaves the current sheet from the outflow region, a thermal particle is inserted at the inflow region of the reconnection zone. Using this approach, electron dynamics and energetics are analyzed in the reconnection region in between the merging flux tubes. We also analyze the effect of the initial spatial distribution of the electrons, and the injection mechanism, which is relevant for future global scale GRRMHD simulations of accretion disks, where particle ensembles are injected in regions showing signs of reconnection.
The particle motion is guided by the electromagnetic fields $\mathbf{E}$ and $\mathbf{B}$ as obtained from the MHD evolution, according to the Lorentz force in CGS units with the speed of light set equal to $c=1$
\begin{equation}
\frac{d\mathbf{u}}{dt} = \frac{q}{m c}\left(\mathbf{E} + \frac{\mathbf{u}\times\mathbf{B}}{\Gamma}\right),
\label{eq:lorentztensCGS}
\end{equation}
where coordinate time $t$ is used as affine parameter, $\Gamma = \sqrt{1 + u^{2}}$ is the particle's Lorentz factor, $\mathbf{u} = \Gamma \mathbf{v}/c$ is the particle four-velocity, divided by the particle's rest mass $m$ and normalized to the speed of light $c=1$.

Reconnection is triggered in a setup of two coalescing Lundquist tubes  (\cite{SironiPorth}; \cite{ripperda2018coalescence}). The flux tubes are given by the magnetic field 
\begin{equation}
\mathbf{B}(r \leq r_j) = \alpha_t c_t  J_1(\alpha_t r) \mathbf{e_{\phi}} + \alpha_t c_t \sqrt{J_0(\alpha_t r)^2 + \frac{C}{(\alpha_t c_t)^2}} \mathbf{e_z},
\label{eq:lundquist}
\end{equation}
where $J_0$ and $J_1$ are Bessel functions of the zeroth and first kind respectively and the constant $C=0.01$ is chosen such that the minimum $B_z$ component remains positive, the constant $\alpha_t \approx 3.8317$ is the first root of $J_0$, and $c_t = 0.262$ ensures that the maximum value of the $B_z$ field is unity, corresponding to the central value in the flux tubes. The solution is terminated at $r = r_j = 1$, corresponding to the first zero of $J_1$, such that $B_z = B_z(r_j)$ and $B_{\phi} = 0$ for $r > r_j$. The total current in the flux tube is zero and no surface currents are present in the initial setup. The two flux tubes are just touching each other and the centre positions are set at $\mathbf{x_c} = (\pm r_j,0,0)$. The setup is perturbed by a velocity perturbation $\mathbf{v}_{kick} = (\pm 0.1c,0,0)$ that pushes the tubes together, with the $\pm$ corresponding to the left and right rope respectively. The kick is applied via the initial electric field as $\mathbf{E} = -\mathbf{v} \times \mathbf{B}$.
The dynamical evolution is independent of the kick velocity, such that the choice of $\mathbf{v}_{kick}$ is not restricting the validity of the conclusions (\cite{SironiPorth}). For the underlying SRRMHD results, as well as the reconnection and plasmoid properties we refer to \cite{ripperda2018coalescence}; For a comparison to PiC results we refer to \cite{SironiPorth}.
The pressure $p_0$ and density $\rho_0$ profiles are initially uniform in the force-free equilibrium setup. The values of $\rho_0$ and $p_0$ are varied between cases to set the magnetization $\sigma_0 = B_0^2 / (\rho_0 h_0)$ and plasma-$\beta_0 = 2 p_0 / B_0^2$, where $h_0 = 1 + 4 p_0 / \rho_0$ is the initial enthalpy density where we assume an ideal gas equation of state with adiabatic index $\hat{\gamma} = 4/3$.

We set the typical length to $L = r_j = 10^{10}$ cm, such that the whole domain is within $x \in [-3L, 3L]$, $y \in [-3L, 3L]$. In all high-resolution ($8192^2$ cells) MHD runs of \cite{ripperda2018coalescence} we initialize 100.000 electrons. 
All particles are initialized from a Maxwellian velocity distribution with temperature $T_0 = 5\times10^{-3} \beta_0 / \rho_0$, thermal velocity $v_{th} = \sqrt{T_0}$ and thermal Lorentz factor $\Gamma_{th} \approx 1$, in accordance with the MHD temperature as set by $\beta_0 = 2 p_0 / B^2$ and $\sigma_0 = B^2 / (\rho_0 h_0)$. We explore a range of uniform resistivities $\eta = 10^{-2}$, $\eta = 10^{-3}$, $\eta = 10^{-4}$, $\eta = 5\times10^{-5}$ and $\eta = 0$, resulting in Lundquist numbers $S = 1/\eta$ of 100, 1000, 10000, 20000 and $\infty$, where \cite{ripperda2018coalescence} found a critical Lundquist number of $S_c \geq 20000$ for plasmoid formation to occur in SRRMHD. We also explore a non-uniform resistivity model $\eta(r,t) = \eta_0(1+\Delta^2_{ei} J)$ \cite{lingam2017}, depending on the current density magnitude $J$ and the asymptotically small parameter $\Delta^2_{ei}$ that is varied between cases. Two cases with varying plasma-$\beta_0$ of 0.1 and 0.5 are considered and three cases with varying magnetization $\sigma_0$ of 0.9, 1.0, and 3.3 are explored by varying the pressure $p_0$ and the density $\rho_0$ in a regime that is particularly relevant for reconnection in black hole accretion disks \cite{ball2017}. The electromagnetic fields are obtained from SRRMHD simulations, except for run Gi, where the setup is evolved with ideal SRMHD (i.e. $\eta = 0$).
\begin{table*}
	\centering
	\caption{The simulated cases with characteristic parameters: kick velocity $v_{kick,x}$; plasma-$\beta_0 = 2p_0/B_0^2$; magnetization $\sigma_0 = B^2_0 / (\rho_0 h_0)$; base resistivity $\eta_0$; nonuniform resistivity factor $\Delta_{ei}^{2}$; and maximum resistivity in the domain over the total time evolution $\eta_{max}$.}
	\label{table:allruns}
	\scriptsize
	\begin{tabular}{lcccccccccccr}
		\hline
		Run & $v_{kick,x}$ & $\beta_0$ & $\sigma_0$ &  $\eta_0$ & $\Delta^2_{ei}$   & $\eta_{max}$ \\
		\hline
		A & $0.1c$ & 0.1 & 3.3& $5 \times 10^{-5}$ & 0 & $5 \times 10^{-5}$  \\
		B  & $0.1c$ & 0.1 & 3.3& $1\times 10^{-4}$ & 1 &$404.7\times10^{-4}$  \\
		C  & $0.1c$ & 0.1 & 3.3& $1\times10^{-4}$ & 0.1 & $264.6\times10^{-4}$  \\
		D  & $0.1c$ & 0.1 & 3.3& $1\times10^{-4}$ & 0.01 &$6.7\times10^{-4}$  \\
		E  & $0.1c$ & 0.1 & 3.3& $1\times 10^{-4}$ & 0.001 & $1.3\times10^{-4}$  \\
		F   & $0.1c$ & 0.1 & 3.3& $1 \times 10^{-4}$  & 0 &$1 \times 10^{-4}$   \\
		G   & $0$ & 0.1 & 3.3& $1 \times 10^{-4}$  & 0 &$1 \times 10^{-4}$    \\
                 Gi    & $0$ & 0.1 & 3.3& $0$  & 0 &$0$  \\
	        H   & $0.1c$ & 0.1 & 3.3& $1 \times 10^{-3}$  & 0 &$1 \times 10^{-3}$ \\
		I    & $0.1c$ & 0.1 & 3.3& $1 \times 10^{-2}$  & 0 &$1 \times 10^{-2}$ \\
		J  & $0.1c$ & 0.5 & 0.9& $5 \times 10^{-5}$  & 0 &$5 \times 10^{-5}$ \\
		K & $0.1c$ & 0.5 & 1.0& $5 \times 10^{-5}$  & 0 &$5 \times 10^{-5}$ \\
		\hline
			\end{tabular}
\end{table*}
The particles are either uniformly initialized in the region of the magnetic flux tubes, $x \in [-2L,2L]$, $y \in [-1L,1L]$, or, in runs indicated by an ``s", particles are initialized in a smaller box around the current sheet with $x \in [-0.1L,0.1L]$, $y \in [-1L,1L]$. 
In the latter case, any particle that leaves this box is replaced by a thermal particle and inserted at a random $y$-position $\in [-1L,1L]$ at $x = \pm 0.05L$. In this way only the particles that accelerate in the formed current sheet are taken into account and bulk acceleration in the initial flux ropes is neglected. The thermal part of the distribution is in this case determined by the injected particles; However, we are mainly interested in the particles that accelerate to non-thermal energies in the reconnection layer.

All runs are summarised in Table \ref{table:allruns}. The first four parameters, the perturbation velocity $v_{kick,x}$, background plasma-$\beta_0$, background magnetization $\sigma_0$ and base resistivity $\eta_0$ are properties of the SRRMHD simulations. A non-zero $\Delta_{ei}$-parameter indicates a nonuniform resistivity run. The maximum value of the resistivity $\eta_{max}$, taken in the whole simulation box and over the whole simulation time, in such cases is given in the last column.

\section{Plasmoid formation and particle acceleration in flux merging events}
In Figure \ref{fig:gammaparticles}, the spatial particle distribution is shown for runs G (left-hand column, $v_{kick,x} = 0$, $\eta = 10^{-4}$, $\beta_0 = 0.1$, $\sigma_0 = 3.3$), F (middle column, $v_{kick,x} = 0.1$, $\eta = 10^{-4}$, $\beta_0 = 0.1$, $\sigma_0 = 3.3$) and A (right-hand column, $v_{kick,x} = 0.1$, $\eta = 5\times10^{-5}$, $\beta_0 = 0.1$, $\sigma_0 = 3.3$) at times $t=5 t_c$, $t=10 t_c$, $t=18 t_c$, $t=24 t_c$.
The particles are colored by their Lorentz factor $\Gamma$, as a proxy for their energy $\mathcal{E} = \Gamma m c^2$. Particles accelerate at early times, in the original flux ropes. They are then drawn towards the current sheet and continue to accelerate in the current sheet in runs F and A. 
The particles are expelled in the reconnection outflow regions at the top and the bottom of the reconnection zone, after they have moved through the current sheet. They then start moving around the flux ropes, guided by the resistive electric field, potentially re-entering the current sheet. 
In run A ($\eta = 5\times10^{-5}$), one can observe plasmoids (at $t_c = 18$ and $t_c=24$), consisting of energetic particles in the current sheet. These plasmoids are expelled in the reconnection outflow regions at the top and bottom of the current sheet.
In unperturbed run G (left-hand column), the flux tubes do not merge and no current sheet forms. Contrary to the PiC results in \cite{SironiPorth}, here we observe particle heating in the flux ropes. This is caused by the relatively high resistivity $\eta = 10^{-4}$, that was chosen to be close to the critical Lundquist number $S_c = 1/\eta \approx 20.000$; where plasmoids form for $S > S_c$. The particles accelerate to moderate Lor{entz factors of $\sim \mathcal{O}(10)$ inside the flux tubes due to the formation of a resistive electric field $\sim \eta \mathbf{J}$. In ideal SRMHD run Gi (that is not shown here), this acceleration does not occur, due to the ideal approximation $\eta=0$. Ideal SRMHD simulations however unavoidably show signs of numerical resistivity, resulting in reconnection. Compared to SRRMHD this effect is uncontrolled, since there is no handle on the resistivity.
\begin{figure*} 
\begin{center}
\subfloat{\includegraphics[width=0.33\columnwidth, clip=true]{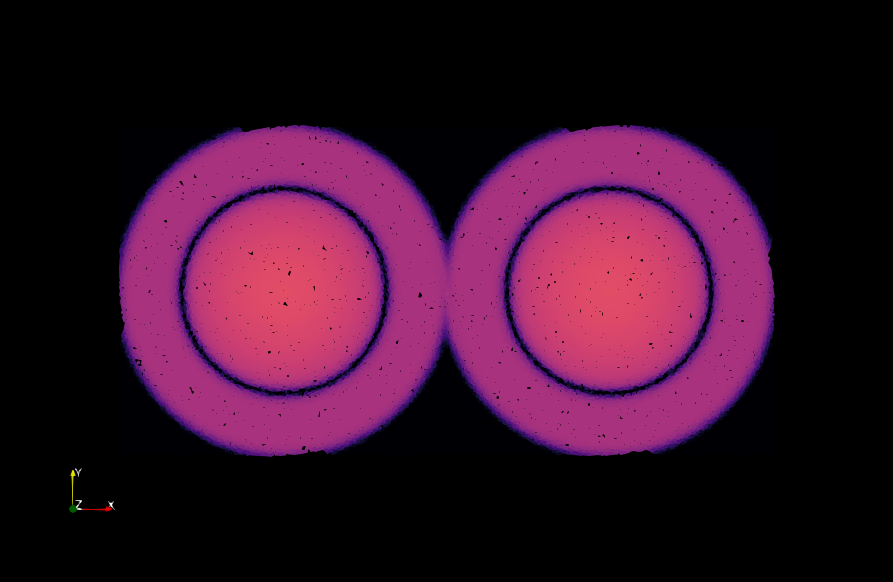}}
\subfloat{\includegraphics[width=0.33\columnwidth, clip=true]{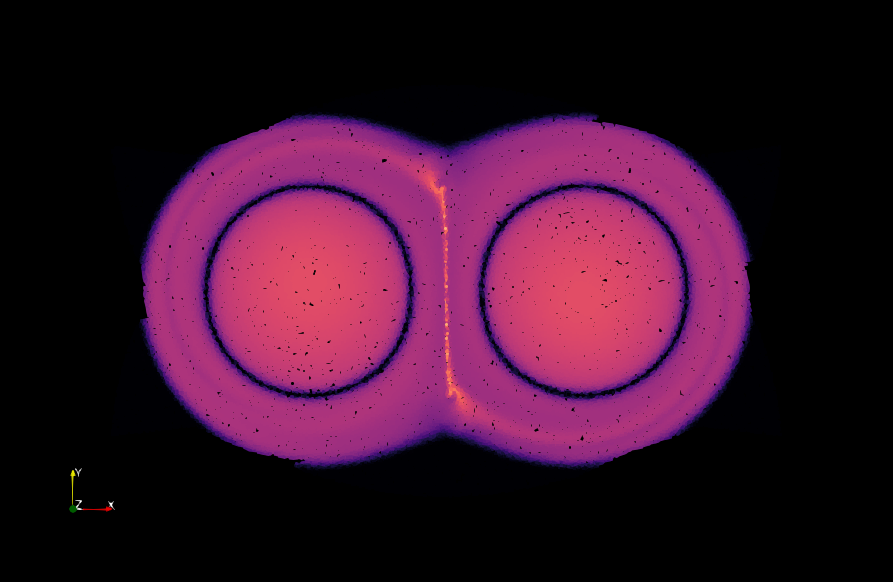}}
\subfloat{\includegraphics[width=0.33\columnwidth, clip=true]{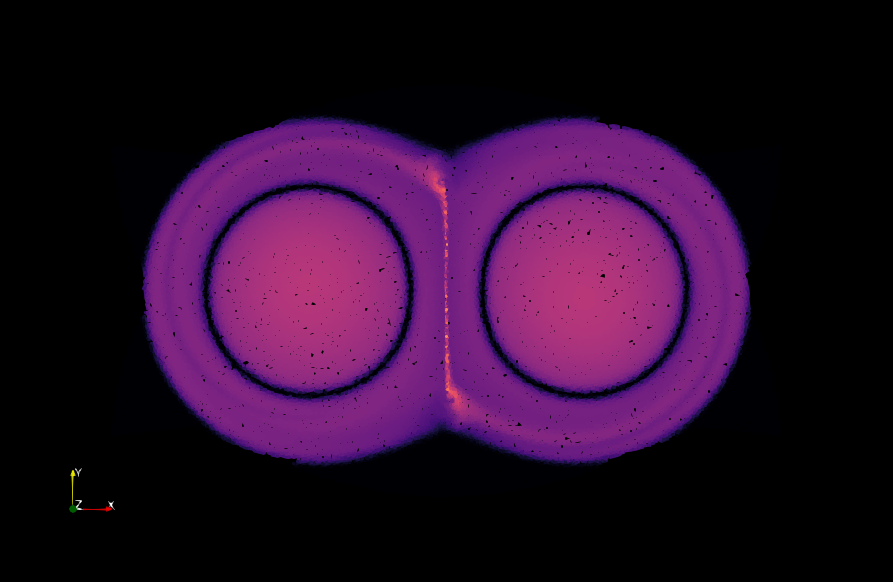}}

\subfloat{\includegraphics[width=0.33\columnwidth,  clip=true]{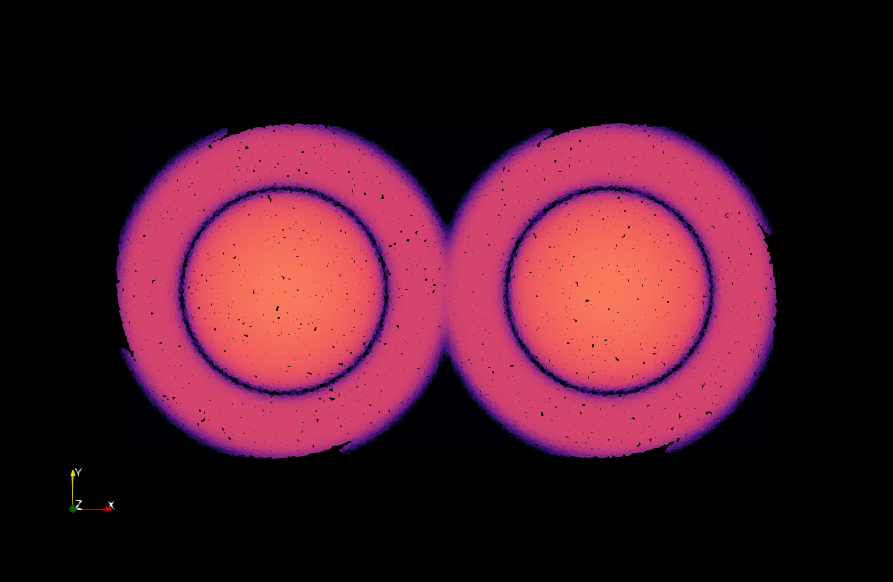}}
\subfloat{\includegraphics[width=0.33\columnwidth, clip=true]{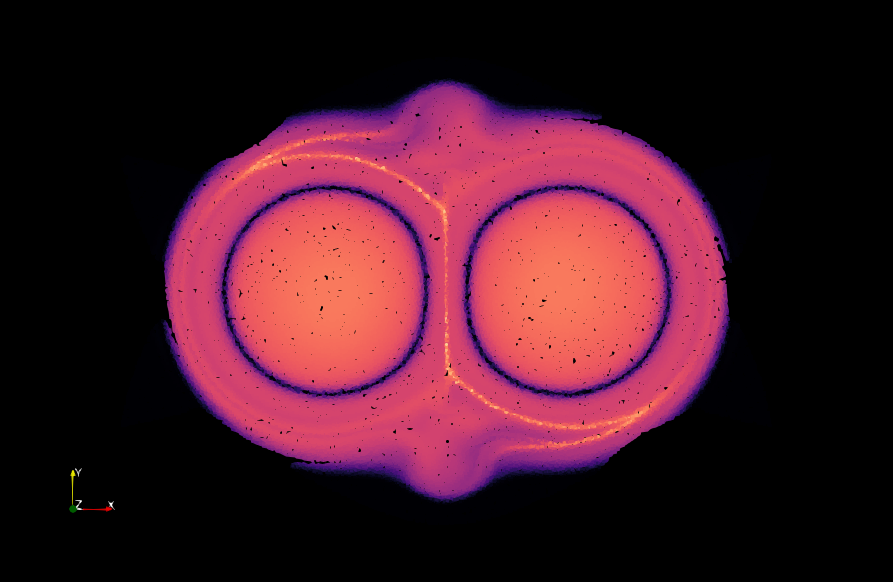}}
\subfloat{\includegraphics[width=0.33\columnwidth, clip=true]{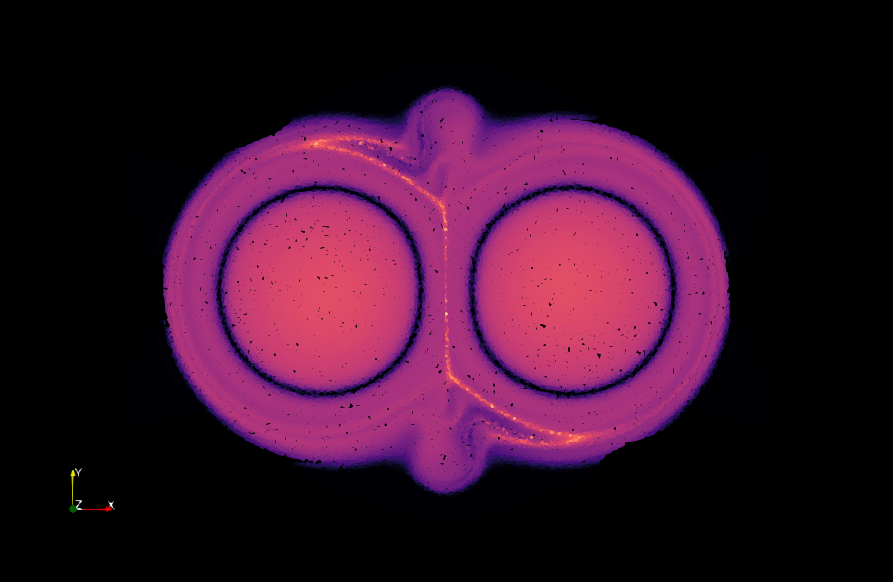}}

\subfloat{\includegraphics[width=0.33\columnwidth,  clip=true]{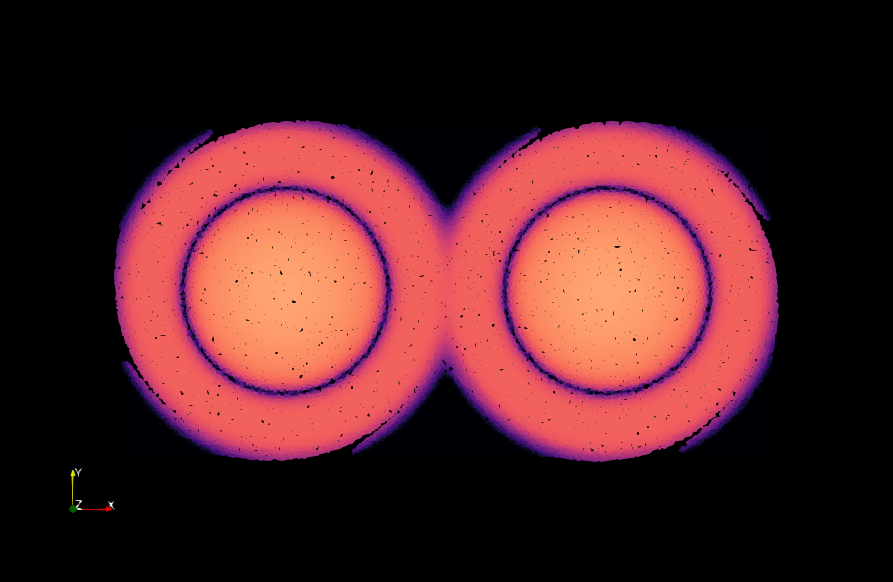}}
\subfloat{\includegraphics[width=0.33\columnwidth,  clip=true]{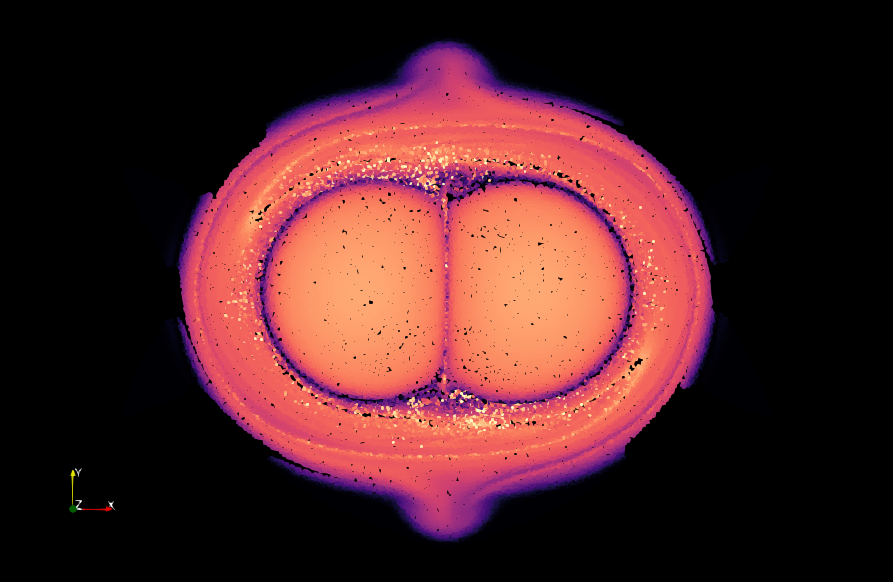}}
\subfloat{\includegraphics[width=0.33\columnwidth,  clip=true]{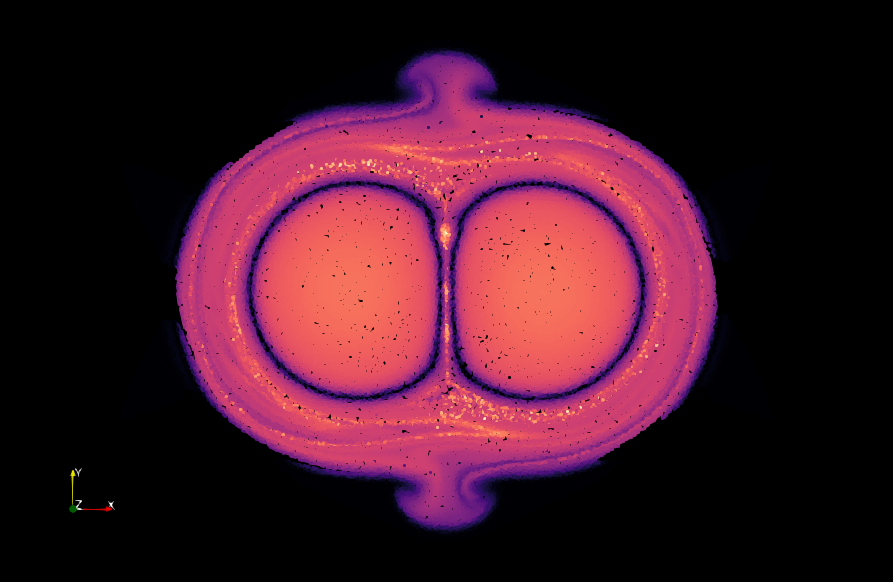}}

\subfloat{\includegraphics[width=0.33\columnwidth,  clip=true]{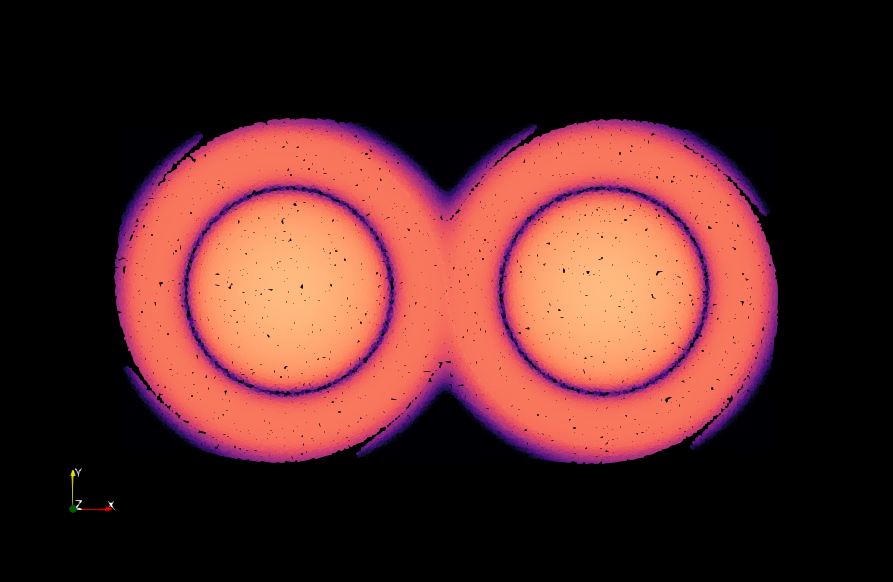}}
\subfloat{\includegraphics[width=0.33\columnwidth,  clip=true]{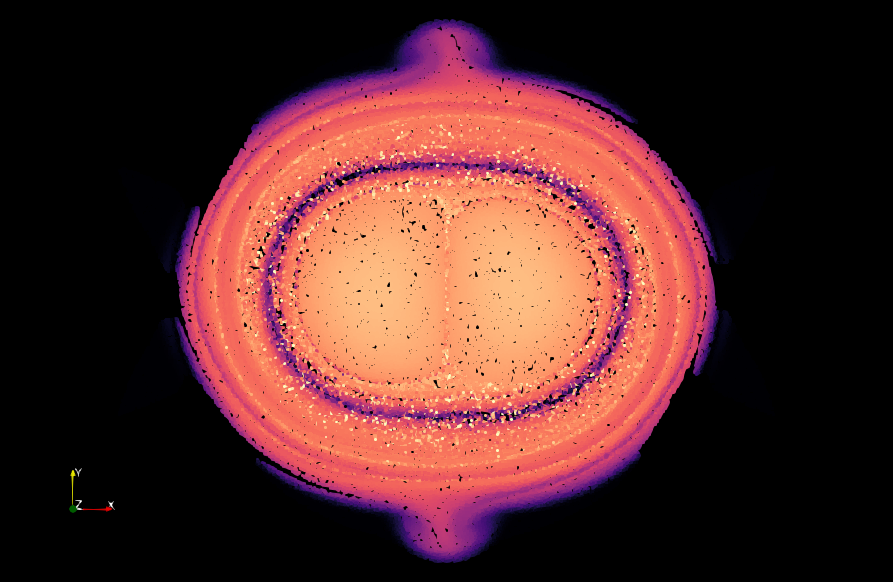}}
\subfloat{\includegraphics[width=0.33\columnwidth,  clip=true]{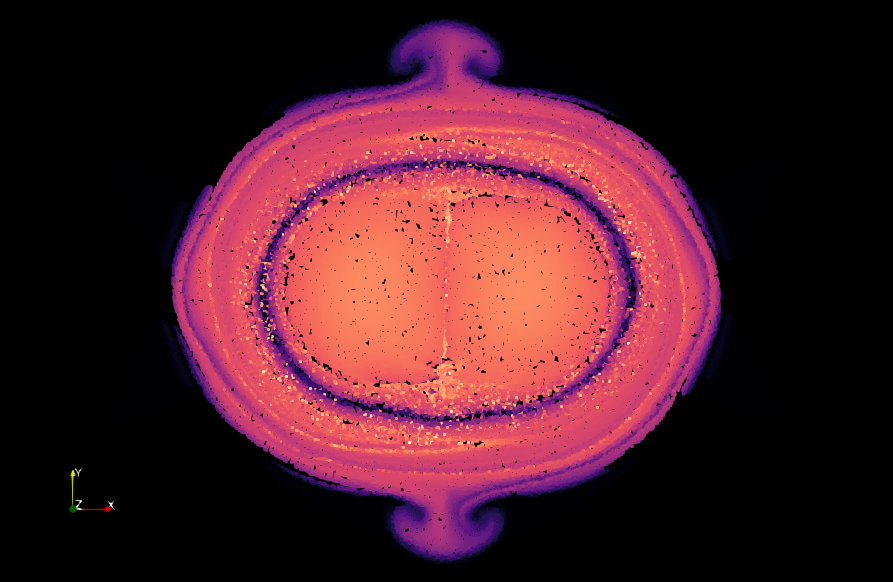}}
\caption{
Particles colored by their Lorentz factor $ \Gamma = (1+u_x^2 + u_y^2 + u_z^2)^{1/2}$ in runs (from left to right) G ($v_{kick,x} = 0$, $\eta = 10^{-4}$, $\beta_0 = 0.1$, $\sigma_0 = 3.33$), F ($v_{kick,x} = \pm 0.1c$, $\eta = 10^{-4}$, $\beta_0 = 0.1$, $\sigma_0 = 3.33$) and A ($v_{kick,x} = \pm 0.1c$, $\eta = 5\times 10^{-5}$, $\beta_0 = 0.1$, $\sigma_0 = 3.33$) at $t=5 t_c$, $t=10 t_c$, $t=18 t_c$, $t=24 t_c$ (from top to bottom respectively). 
The logarithmic color scale is saturated between 1 and 500. 100.000 electrons are initialized with a Maxwellian velocity distribution, randomly distributed in the area covered by the magnetic flux tubes $x \in [-2L,2L]$, $y\in [-1L,1L]$ and evolved alongside the MHD. The particles accelerate in the current sheets and the plasmoids. Plasmoids are visible as a concentration of particles with higher Lorentz factors in the current sheet in between the flux tubes, for case A (right-hand column) at $t=18t_c$.}
\label{fig:gammaparticles}
\end{center}
\end{figure*}

\subsection{Effects of Lundquist number on the energy distribution}
Figure \ref{fig:distributions} shows the distributions for $\Gamma = (1+ u_x^2 +u_y^2 + u_z^2)^{1/2}$, as taken in a box surrounding the current sheet $x \in [-0.05L,0.05L]$, $y \in [-1L,1L]$ for run G (left-hand panels, $v_{kick,x} = 0$, $\eta = 10^{-4}$, $\beta_0 = 0.1$, $\sigma_0 = 3.33$), run F (middle panels, $v_{kick,x} = \pm 0.1c$, $\eta = 10^{-4}$, $\beta_0 = 0.1$, $\sigma_0 = 3.33$) and run A (right-hand panels, $v_{kick,x} = \pm 0.1c$, $\eta = 5\times 10^{-5}$, $\beta_0 = 0.1$, $\sigma_0 = 3.33$). We confirmed the results do not change for a slightly smaller or larger box. In the top panels, particles are initialized in the whole simulation box, including the flux tubes $x \in [-2L,2L]$, $y \in [-1L,1L]$. In the bottom panels, particles are only initialized in the reconnection layer at $x \in [-0.05L,0.05L]$, $y \in [-1L,1L]$, such that the distribution is not affected by particles entering the current sheet after being heated in the flux tubes.

In the unperturbed run G (top left-hand panel), a non-thermal tail, with $\Gamma_{max} \approx 10$ forms at later times ($t \gtrsim 18 t_c$, green and red lines) by particles that are accelerated in the flux tubes and enter the box where the distribution is taken. If the particles are initialized in the box $x \in [-0.05L,0.05L]$, $y \in [-1L,1L]$ in run Gs, no non-thermal tail forms and particles accelerate only mildly to $\Gamma_{max} \approx 3$ due to the resistive electric field $\mathbf{E} \sim \eta \mathbf{J}$ present in between the flux tubes. Note that in these runs, no current sheet has formed.

In cases F (top middle panel) and A (top right-hand panel) a non-thermal tail forms in the current sheet from $t \gtrsim 5 t_c$ onwards. The maximum Lorentz factor is proportional to the resistivity $\Gamma_{max} \approx 10^3$ in case F ($\eta = 10^{-4}$) and $\Gamma_{max} \approx 5\times10^2$ in case A ($\eta = 5\times10^{-5}$). In both cases, a ``bump'' in the energy distribution can be observed at $\Gamma \approx 10^2$ at times $t = 18 t_c$ (green lines) and $t = 24 t_c$ (red lines). This is caused by electrons entering the current sheet after they have been heated in the flux tubes, due to a resistive electric field $\mathbf{E} \sim \eta \mathbf{J}$. This bulk heating in the large magnetic flux tubes is similar to the effect observed in Figure \ref{fig:gammaparticles}. In PiC simulations, such effects are not observed, due to an intrinsically low resistivity in the flux tubes and ambient plasma \cite{SironiPorth}. 

For run Gi (not shown here), in ideal SRMHD ($\eta = 0$), no non-thermal tail forms at all and particles do no accelerate in the flux tubes, confirming that a too large resistivity in the ambient plasma is the cause for the bulk heating. More realistic, non-uniform resistivity models, with a low resistivity in the ambient and a high resistivity in the current sheet, can potentially solve this heating effect in test particles simulations. 

initializing particles in the current sheet box at $x \in [-0.05L,0.05L]$, $y \in [-1L,1L]$ only, reduces the heating effect and results in a power-law distribution for cases Fs (bottom middle panel) and As (bottom right-hand panel). Here, the particles leaving the current sheet are destroyed, and new thermal particles are injected from the reconnection inflow region at random vertical position $y \in [-1L,1L]$ and at $x \pm 0.05L$. The maximum Lorentz factor remains similar for both cases as in the top panels, and is still proportional to the resistivity, but the ``bump'' at $\Gamma \approx 10^2$ disappears. 

In runs A and As (right-hand panels), the plasmoid instability is activated from $t_c \approx 18$ onwards for $\eta = 5\times10^{-5}$, and a change in the distribution is observable in the green and red lines, compared to run F (middle panels) for $\eta=10^{-4}$. It is however unclear if this can be attributed to the particle acceleration in the plasmoids, or due to the larger current density in the current sheet due to the plasmoid instability.

Run Fs has been confirmed to give equivalent distributions with 10.000 electrons and in \cite{Ripperda2} it is shown that for test particles, the total number of particles has little effect on the results, since there is no feedback of the particles on the electromagnetic fields and hence no requirement for a certain number of particles per cell. 
\begin{figure*} 
\begin{center}
\subfloat{\includegraphics[width=0.33\columnwidth]{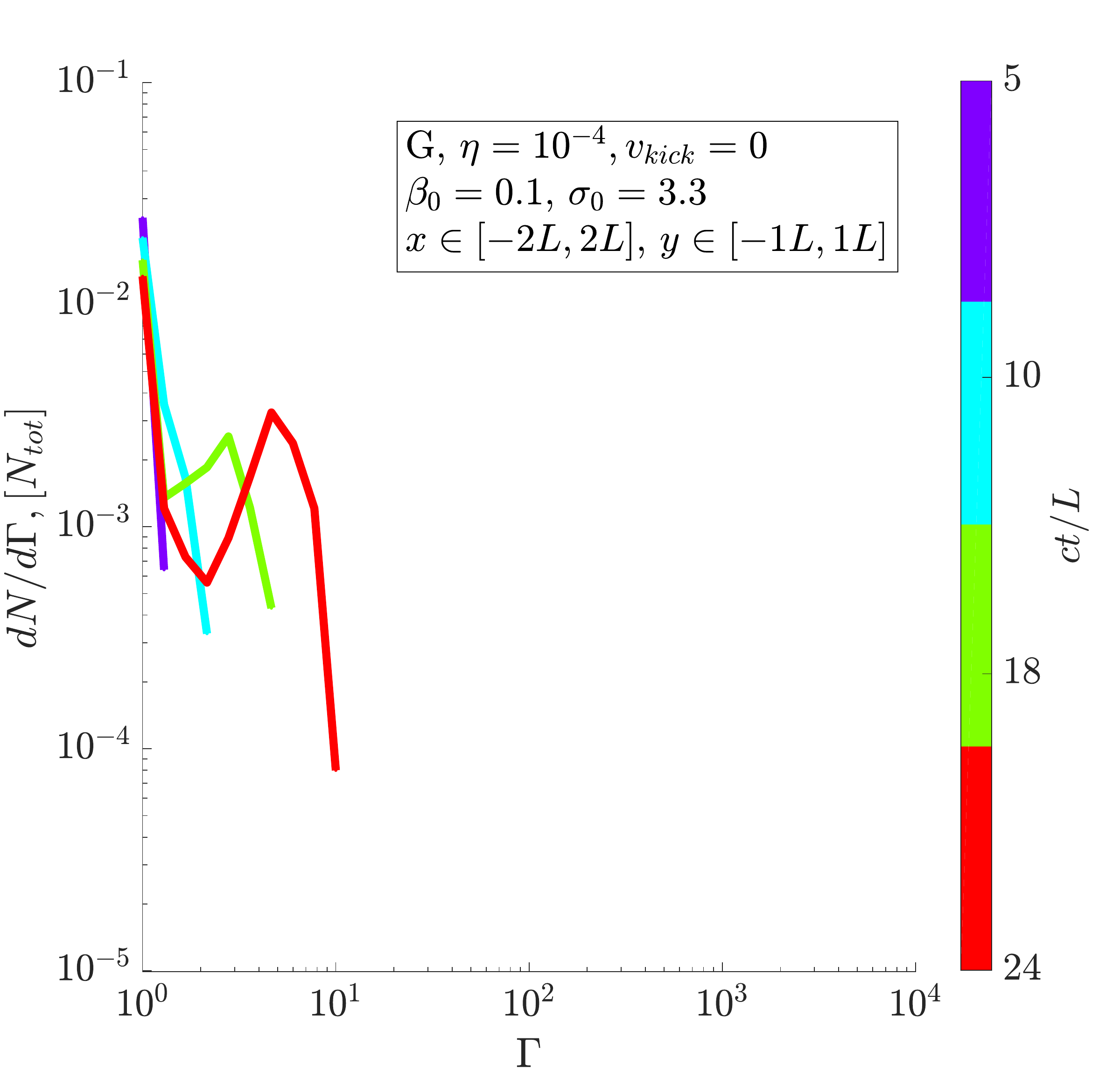}}
\subfloat{\includegraphics[width=0.33\columnwidth]{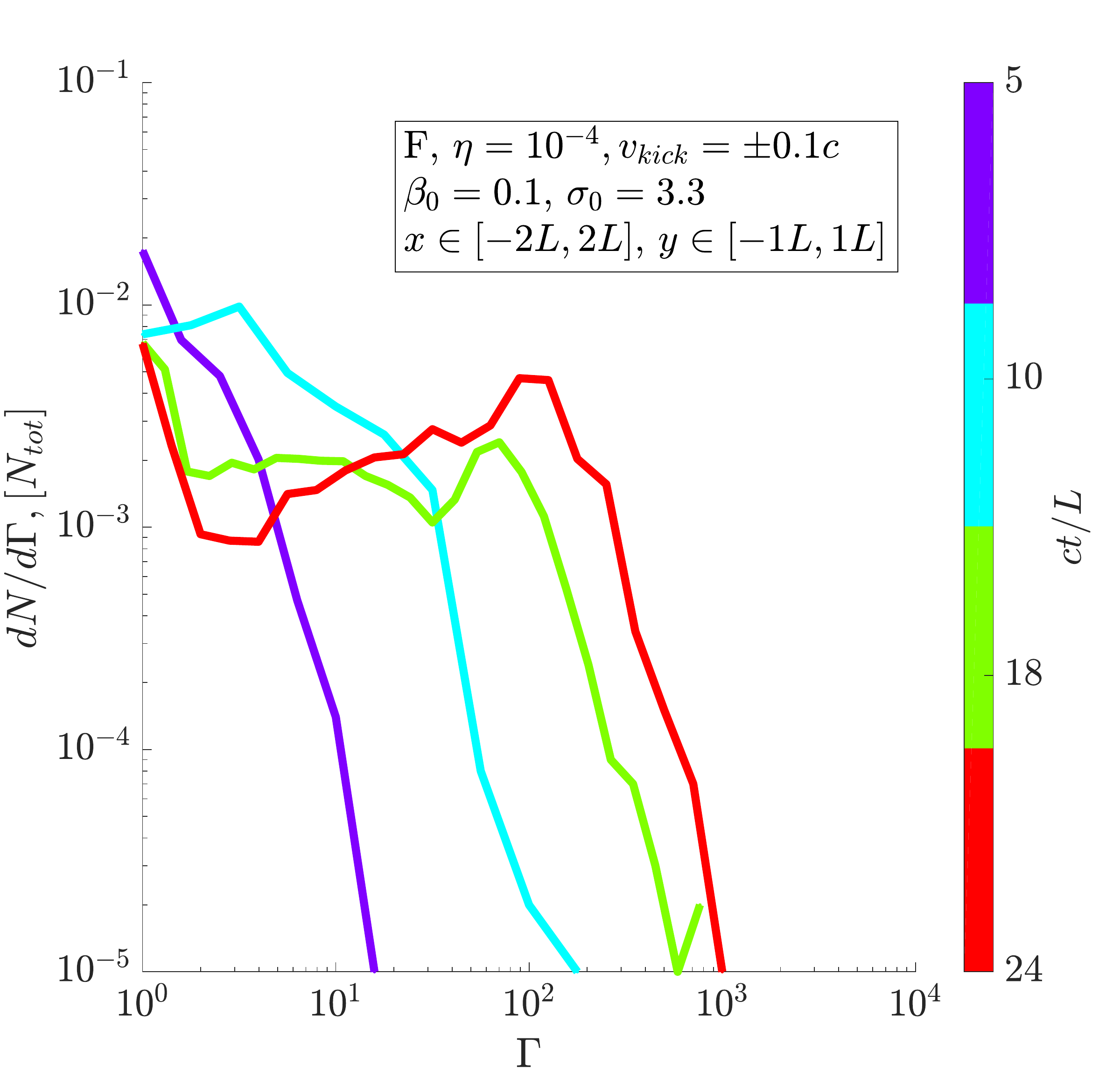}}
\subfloat{\includegraphics[width=0.33\columnwidth]{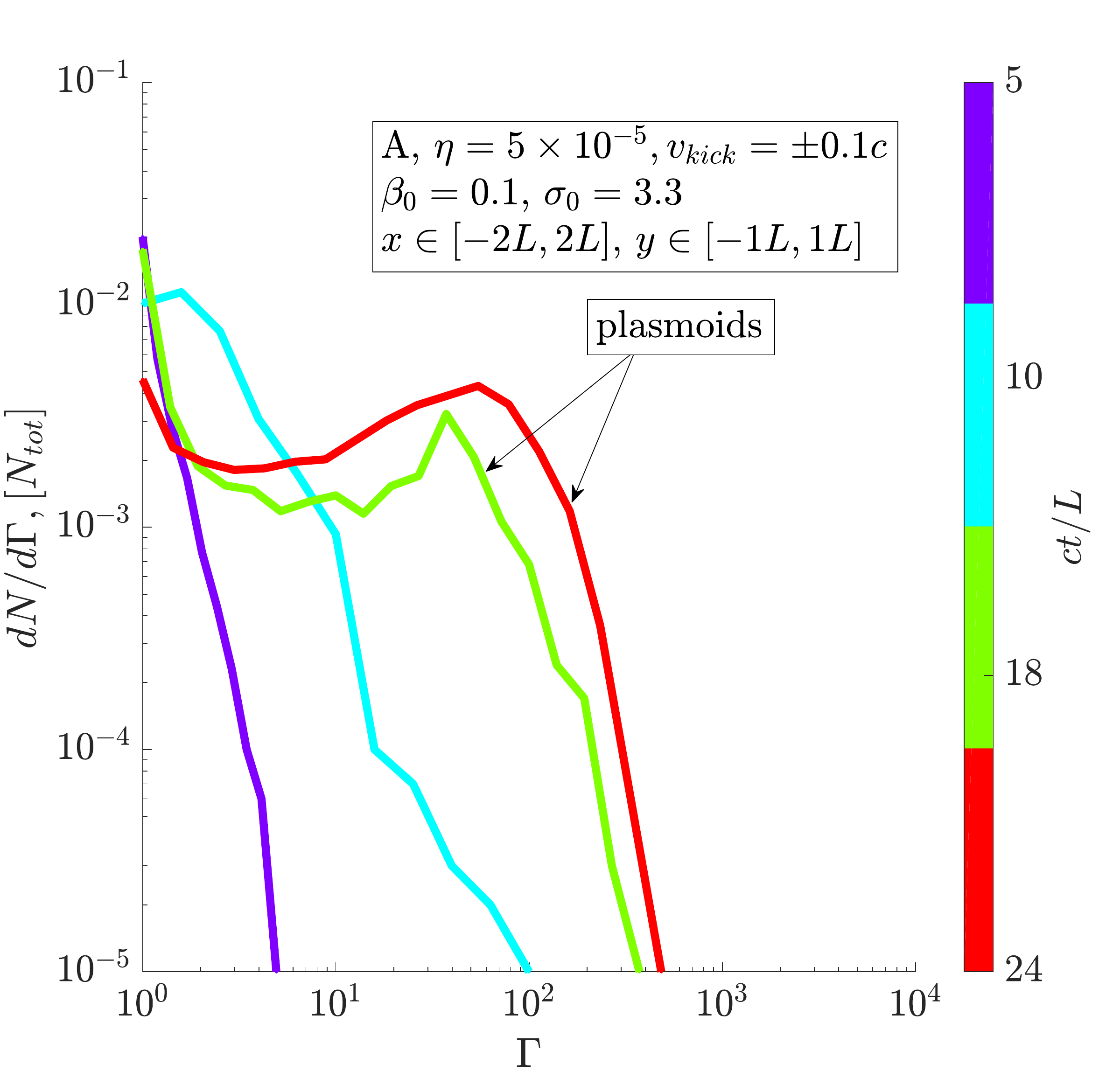}}

\subfloat{\includegraphics[width=0.33\columnwidth]{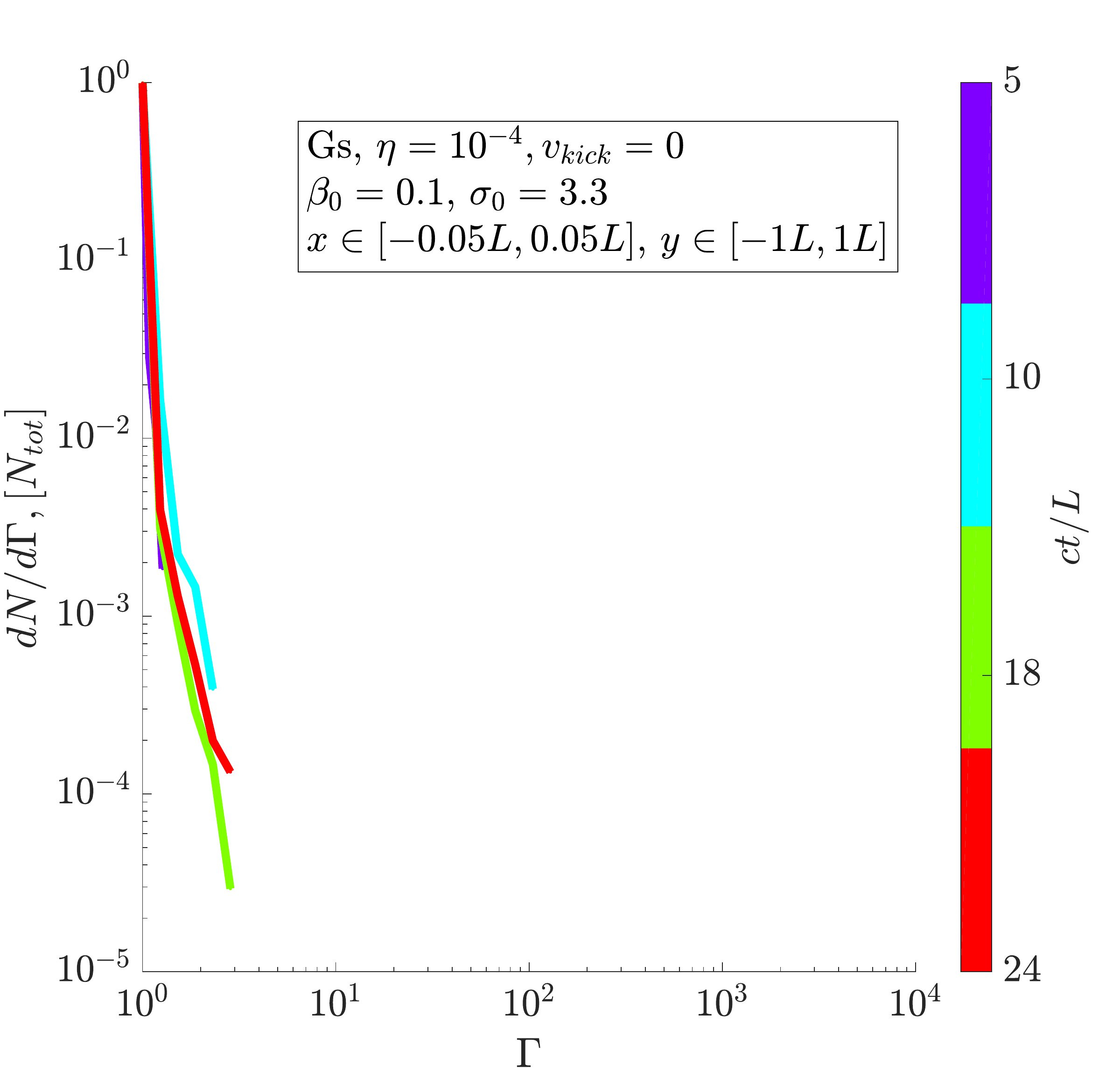}}
\subfloat{\includegraphics[width=0.33\columnwidth]{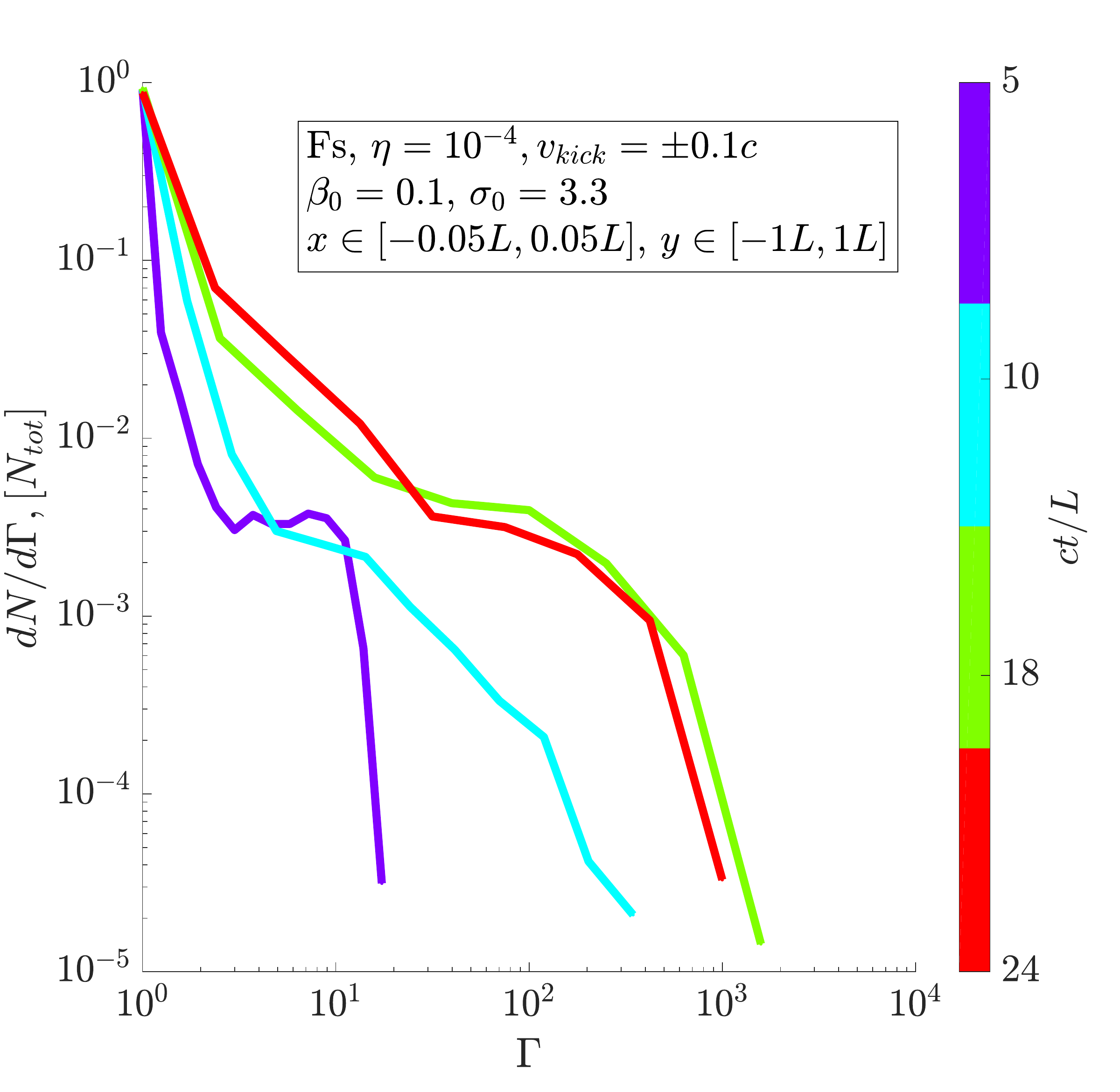}}
\subfloat{\includegraphics[width=0.33\columnwidth]{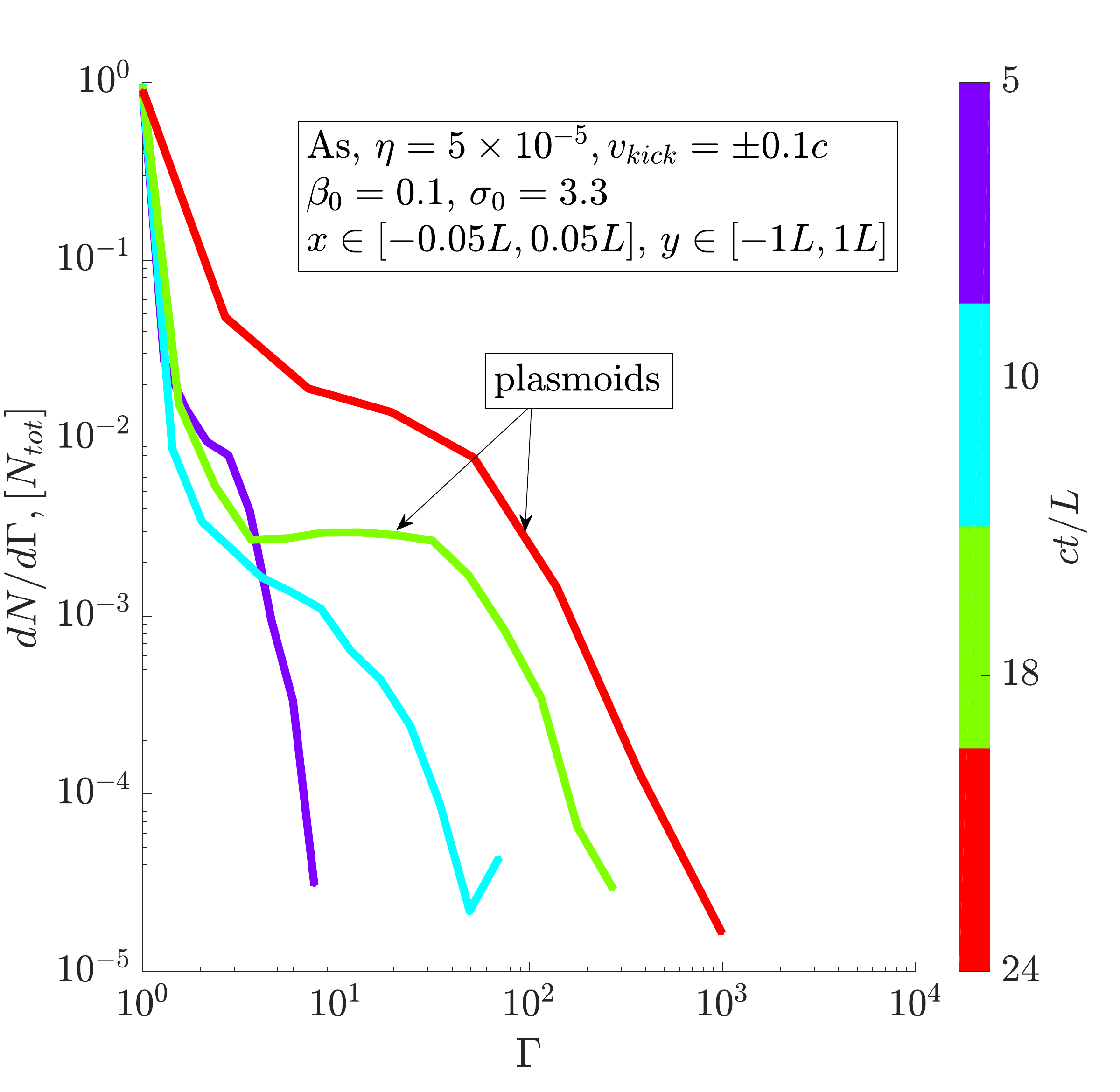}}
\caption{Normalised electron energy distributions ($\mathcal{E}/(mc^2) = \Gamma$) for runs G (top left-hand panel), F (top middle panel), A (top right-hand panel), Gs (bottom left-hand panel), Fs (bottom middle panel) and As (bottom right-hand panel) at times $t=5 t_c$, $t = 10t_c$, $t = 18 t_c$ and $t=24 t_c$ as indicated by the color bar. The distributions are taken in a box $x \in [-0.05, 0.05]$, $y \in [-1,1]$ around the current sheet. In the bottom panels, for runs with indicator ``s'', particles are initialized in the current sheet and particles leaving the reconnection zone are destroyed. This results in a power-law distribution without the ``bump'' in energy at $\Gamma \approx 10^2$ as in the top panels where the particles are initialized in the whole simulation box. The heating in the flux tubes causing the bump is attributed to the large resistivity in these regions.}
\label{fig:distributions}
\end{center}
\end{figure*} 

\subsection{Effects of nonuniform resistivity on the energy distribution}
Figure \ref{fig:distributions_ar_betasigma} shows the Lorentz factor distributions for, from left to right, case Cs (nonuniform resistivity $\eta = 10^{-4}(1+0.1J)$, $\beta_0 = 0.1$, $\sigma_0 = 3.3$), case Js (uniform resistivity $\eta = 5\times10^{-5}$, $\beta_0=0.5$, $\sigma_0=0.9$) and case Ks (uniform resistivity $\eta = 5\times10^{-5}$, $\beta_0=0.5$, $\sigma_0=1.0$). Distributions are again taken in the box around the current sheet at $x \in [-0.05L,0.05L]$, $y \in [-1L, 1L]$ and particles are initialized in the same box and destroyed once they leave this box through the reconnection outflows. 

Run Cs (left-hand panel), with nonuniform resistivity $\eta(r,t) = \eta_0(1+\Delta_{ei}^2 J)$, with $\eta_0 = 10^{-4}$ and $\Delta_{ei}^2=0.1$, should be compared to run Fs with uniform $\eta = 10^{-4}$ (and hence $\Delta_{ei}^2=0$) in the bottom middle panel of Figure \ref{fig:gammaparticles}. The nonuniform resistivity in case Cs is close to $\eta = 10^{-4}$ in the ambient plasma and the flux tubes, where the current density remains small. In the current sheet, the resistivity is strongly enhanced due to the large current density to reach a maximum of $\eta_{max} \approx 264.6 \times 10^{-4}$\cite{ripperda2018coalescence}. Due to the strongly enhanced resistivity the MHD evolution is faster, the current sheet broadens and the resistive electric field increases such that particles accelerate at earlier times $t = 10 t_c$ (cyan line) to reach a Lorentz factor $\Gamma_{max} \approx 5 \times 10^3$, approximately 20 times larger than in case Fs. At later times $t \gtrsim 18 t_c$, the current density dissipates due to the enhanced resistivity, the resistive electric field $\mathbf{E} \sim \eta \mathbf{J}$ decreases and particles decelerate again (green and red lines).

\subsection{Effects of magnetization on the energy distribution}
The middle panel of Figure \ref{fig:distributions_ar_betasigma} shows the effect of an increased plasma-$\beta_0 = 0.5$ in run Js (uniform $\eta = 5\times10^{-5}$), compared to run As ($\beta_0 = 0.1$, $\eta = 5\times10^{-5}$) in the bottom right-hand panel of Figure \ref{fig:gammaparticles}. The increase in $\beta_0$ for run Js results in an effectively decreased magnetization $\sigma_0 = B^2 / (\rho_0 h_0) = B^2 / (\rho_0 + 2 B^2 \beta_0) = 0.9$, compared to $\sigma_0 = 3.3$ for run As. From $t \gtrsim 10t_c$ onwards, a non-thermal distributions forms in the current sheet, that grows to a maximum Lorentz factor $\Gamma_{max} \approx 4 \times 10^2$, approximately 2.5 times smaller than in run As. In run Js, no plasmoids are formed in the current sheet, due to the smaller magnetization than in run As.

In run Ks, in the right-hand panel of \ref{fig:distributions_ar_betasigma}, the magnetization is increased to $\sigma_0 = 1.0$ compared to run Js (middle panel), for $\beta_0 = 0.5$ and uniform resistivity $\eta = 5\times 10^{-5}$. Due to the slightly enhanced magnetization, plasmoids are forming now in the current sheet from $t = 20 t_c$ onwards \cite{ripperda2018coalescence}. Comparing red line at $t = 24 t_c$ in the middle panel (Js, $\sigma_0=0.9$, no plasmoid formation) to the red line in the right-hand panel (Ks, $\sigma_0=1.0$, plasmoid formation), we observe that the distribution has a steeper power-law for case Ks. The maximum Lorentz factor $\Gamma_{max} \approx 4 \times 10^2$ is equivalent to run Js, where no plasmoids have formed.
\begin{figure*} 
\begin{center}
\subfloat{\includegraphics[width=0.33\columnwidth]{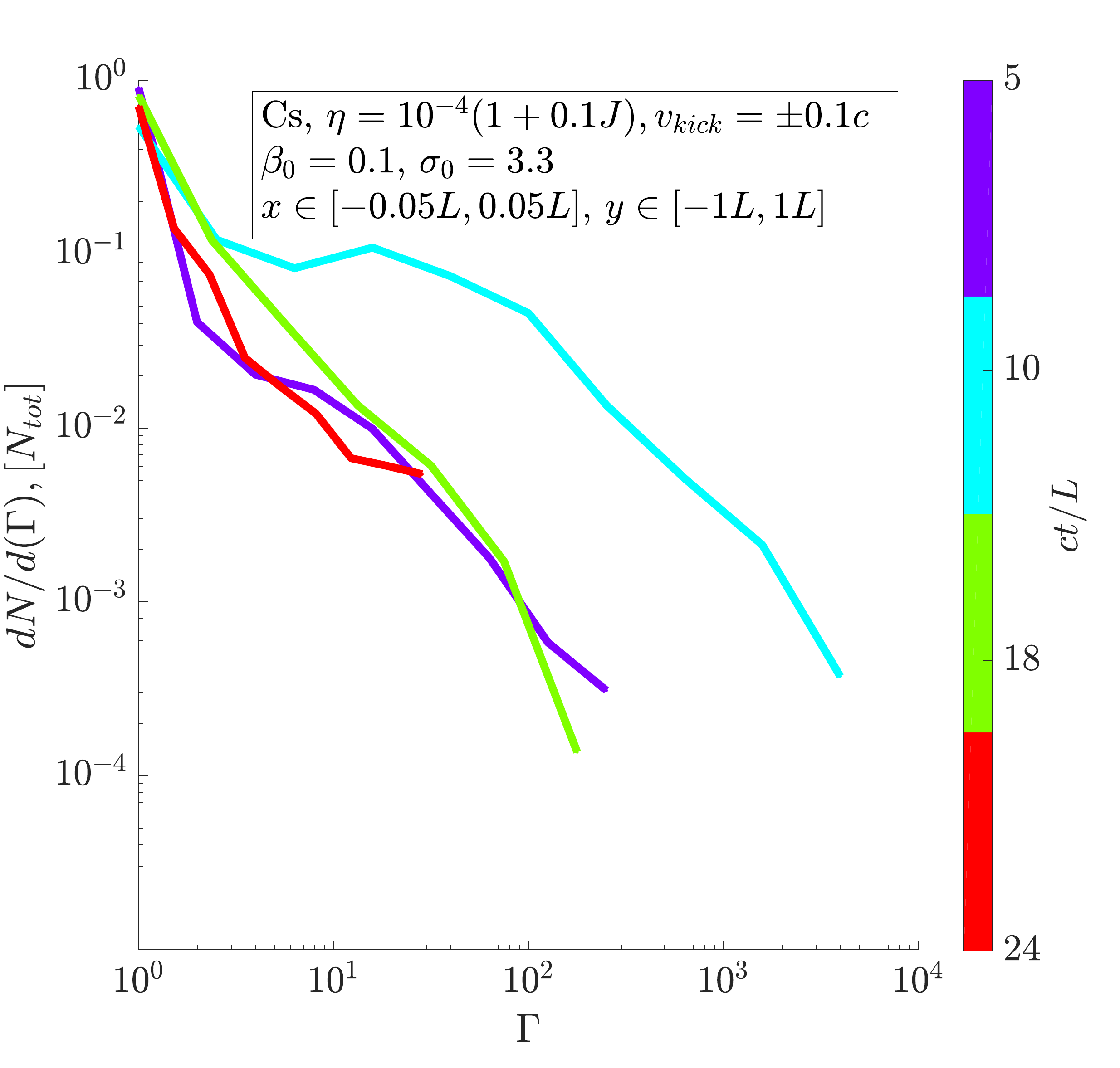}}
\subfloat{\includegraphics[width=0.33\columnwidth]{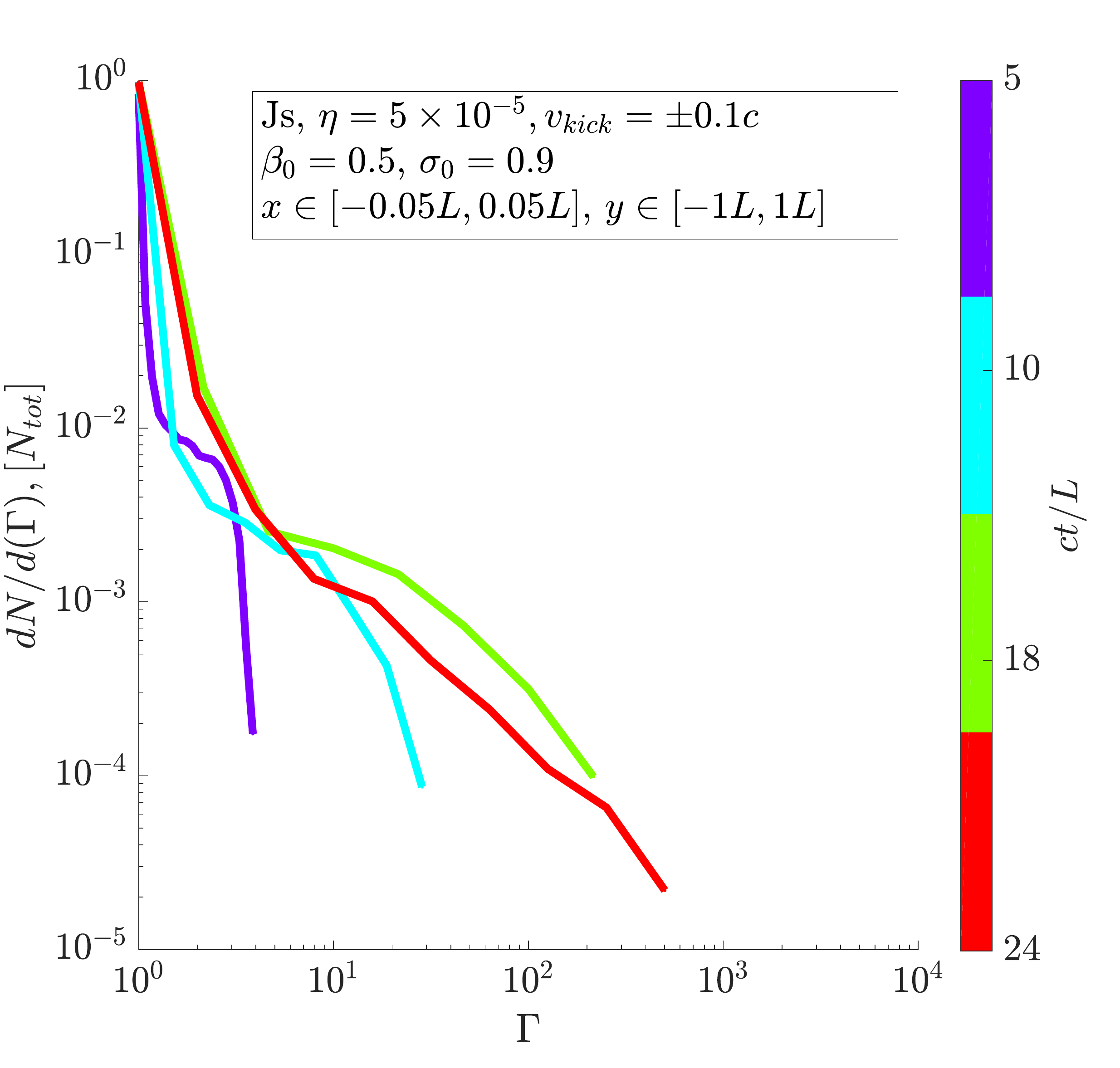}}
\subfloat{\includegraphics[width=0.33\columnwidth]{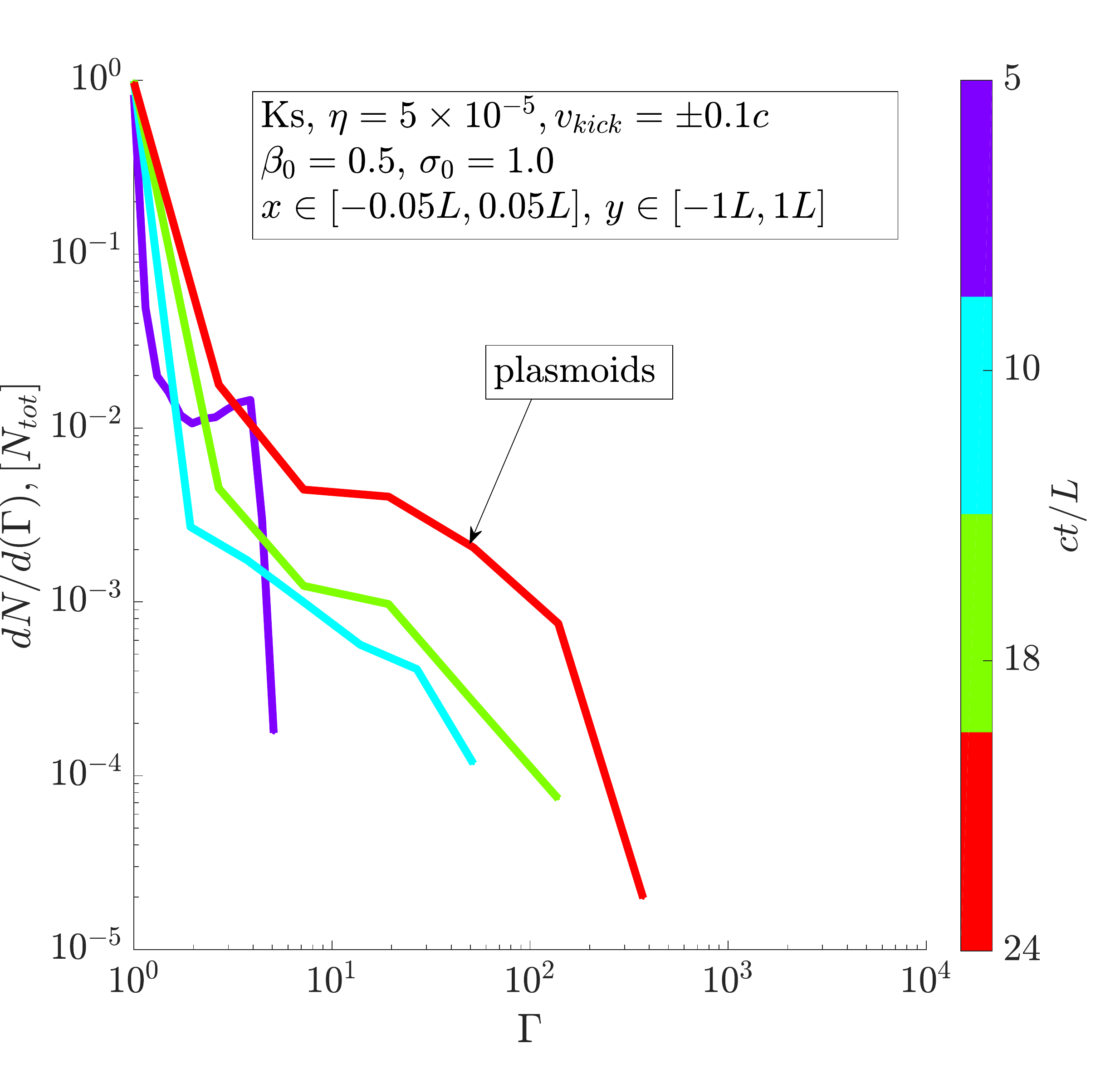}}
\caption{Normalised electron energy distributions ($\mathcal{E}/(mc^2) = \Gamma$) for runs Cs (left-hand panel), Js (middle panel) and Ks (right-hand panel) at times $t=5 t_c$, $t = 10t_c$, $t = 18 t_c$ and $t=24 t_c$ as indicated by the color bar. The distributions are taken in a box $x \in [-0.05, 0.05]$, $y \in [-1,1]$ around the current sheet. Particles are initialized in the current sheet and particles leaving the reconnection zone are destroyed. The left-hand panel shows the effect of a nonuniform resistivity $\eta(r,t) = 10^{-4}(1+0.1 J)$ in run Cs, compared to uniform resistivity $\eta=10^{-4}$ run Fs (bottom middle panel in Figure \ref{fig:distributions}). The middle panel shows the effect of an increased $\beta_0=0.5$ and decreased $\sigma_0=0.9$ for uniform resistivity $\eta = 5\times10^{-5}$ in run Js, compared to run As ($\beta_0=0.1$, $\sigma_0=3.3$, bottom right-hand panel in Figure \ref{fig:distributions}). The right-hand panel shows the effect of plasmoids in run Ks, due to a slightly enlarged $\sigma_0 = 1.0$, compared to run Js.}
\label{fig:distributions_ar_betasigma}
\end{center}
\end{figure*}

\subsection{Lorentz factor evolution}
In Figure \ref{fig:gammavst}, we show the temporal evolution of the maximum Lorentz factor $\Gamma_{max}$ in all runs, taken over the whole ensemble of particles (i.e. not just in the current sheet). A distinction is made between runs with particles in the whole domain (including the initial flux ropes) and runs where particles that end up outside of the current sheet are neglected and destroyed (indicated by index ``s"). In the left-hand panel, all runs with $\beta_0=0.1$, $\sigma_0=3.3$ and uniform resistivity, are compared, with $\eta=5\times10^{-5}$ for run A and As, $\eta=10^{-4}$ for runs F and Fs, $\eta = 10^{-3}$ for run Hs, and $\eta=10^{-2}$ for run Is. Runs G and Gs, with $\eta=10^{-4}$ but without an initial perturbation (i.e. $\mathbf{v}_{kick}=0$) are also shown, as well as run Gis that is evolved with the ideal SRMHD module in {\tt BHAC} (i.e. $\eta=0$). 

All resistive runs show an initial growth of the particle Lorentz factor, that is comparable to the Alfv\'{e}nic growth of the electric field \cite{ripperda2018coalescence}. The acceleration increases for larger resistivity.  In runs Hs (dashed black line) and Is (dashed magenta line), the resistivity is so large that the forming current sheet rapidly diffuses, resulting in a decrease in the maximum Lorentz factor at $t_c \approx 10$ and $t_c \approx 20$, respectively. For runs F (solid green line), Fs (dashed green line), A (solid brown line) and As (dashed brown line), there is a subsequent secondary acceleration phase with larger variability after $t_c \approx 5$. These particles are accelerated by the resistive electric field in the current sheet. Runs F and Fs show that $\Gamma_{max}$ is about two times larger than for runs As and A, in accordance with the factor two difference in resistivity. In runs As and A the plasmoid instability is triggered, which however shows no observable effect on the maximum Lorentz factor. Destroying and injecting particles induces a larger variability for runs As and Fs in the nonlinear regime at $t_c \gtrsim 5$ compared to runs A and F. The maximum Lorentz factor does however not differ much, showing that the particles attain the highest energies inside the current sheet. This confirms the finding that particles accelerate to medium Lorentz factors in the initial flux ropes. 
\begin{figure*} 
\begin{center}
\includegraphics[width=0.32\columnwidth]{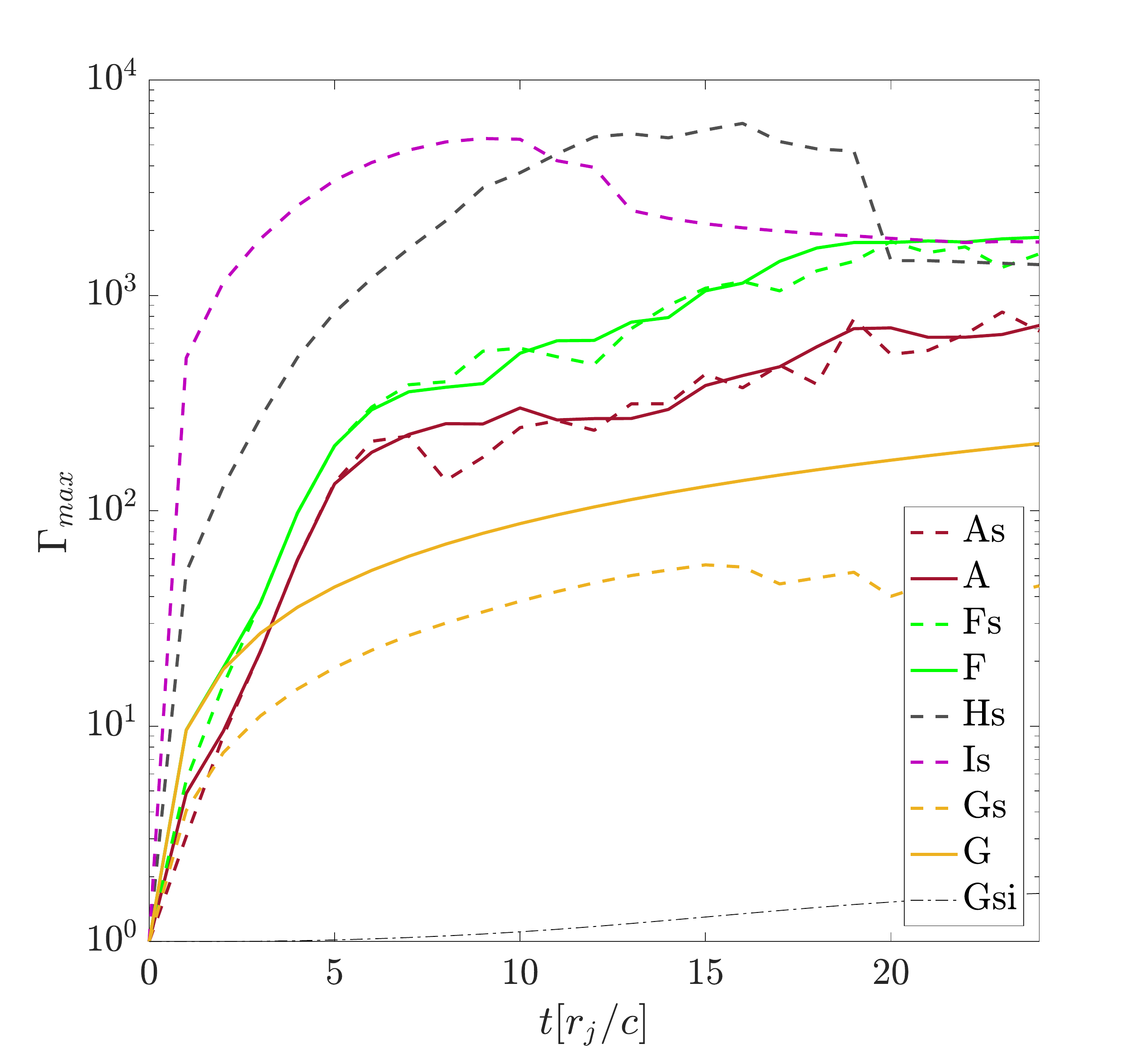}
\includegraphics[width=0.32\columnwidth]{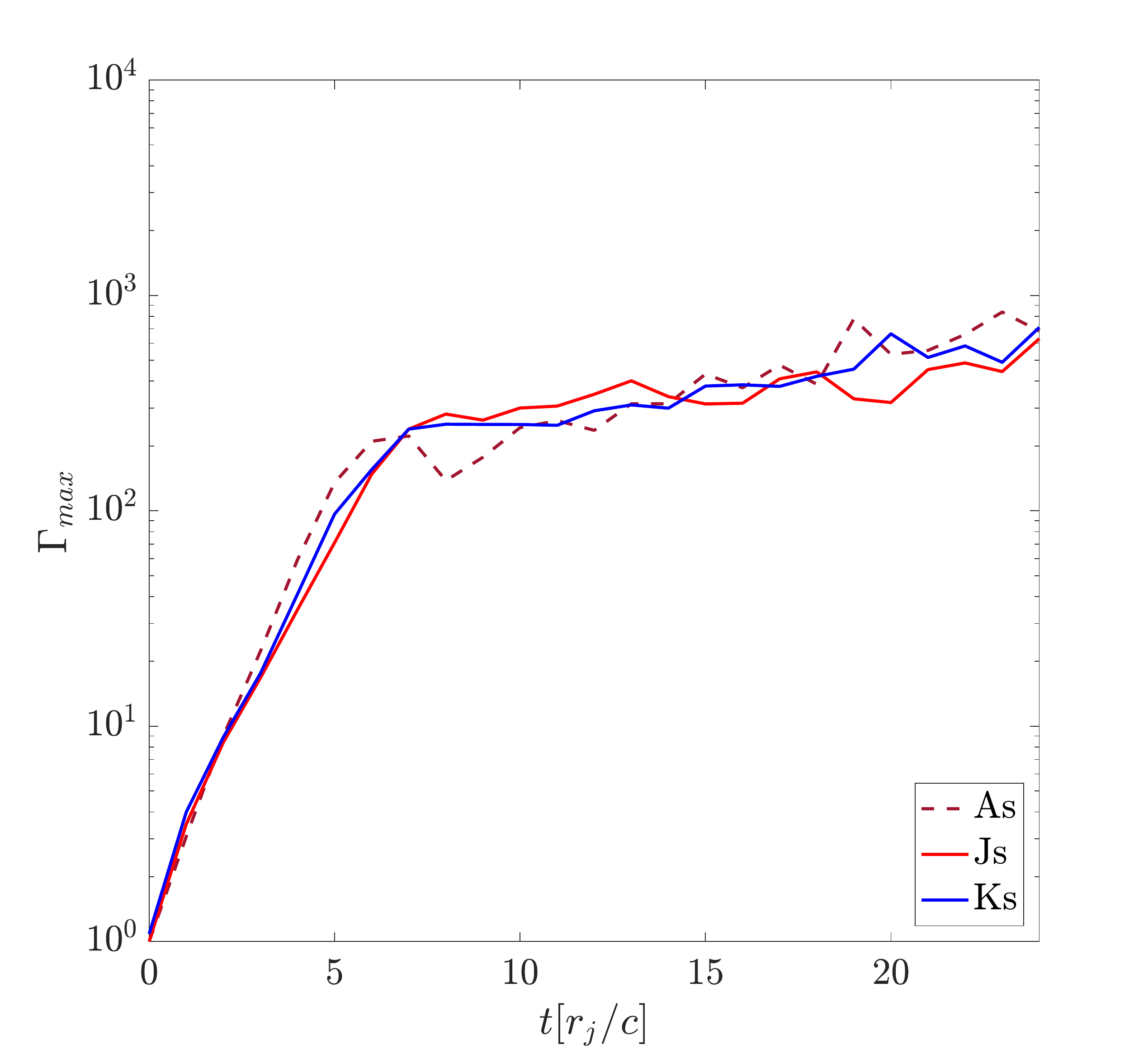}
\includegraphics[width=0.32\columnwidth]{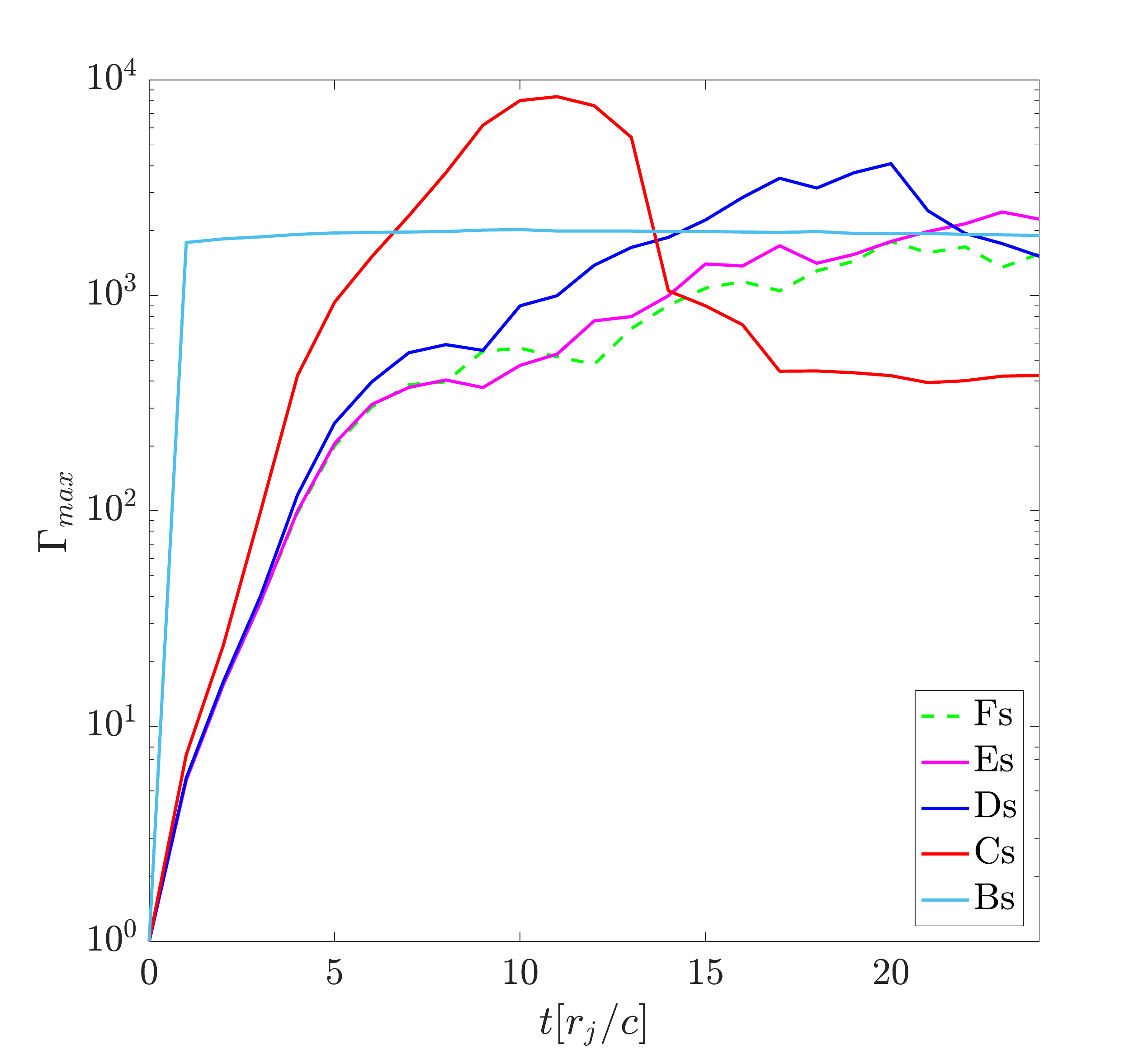}
\caption{Temporal evolution of $ \Gamma_{max}$ where $max( \Gamma)$ is taken in the whole domain, over all particles.  
Runs A, As, F, Fs, Hs, Is, G, Gs and Gsi are shown in the left-hand panel, where runs As and A have resistivity $\eta = 5 \times 10^{-5}$, F and Fs have $\eta = 10^{-4}$, Hs has $\eta = 10^{-3}$, and Is has $\eta = 10^{-2}$. Runs G and Gs remain unperturbed (i.e. $\mathbf{v}_{kick} = 0$) and have $\eta = 10^{-4}$ and run Gsi is both unperturbed and evolved in ideal SRMHD (i.e. $\eta = 0$ by definition). An ``s" in the index (all dashed lines in the left-hand panel) indicates that in these runs particles are removed when they leave the current sheet at $x \in [-0.1L,0.1L]$, $y \in [-1L,1L]$ and in their place a thermal particle is injected at random vertical position $y \in [-1L,1L]$ and at $x \pm 0.05L$. In the middle panel, all runs with $\eta = 5\times10^{-5}$ are shown, where run As has $\beta_0=0.1$, $\sigma_0=3.3$, run Js has $\beta_0=0.5$, $\sigma_0=0.9$ and run Ks has $\beta_0=0.5$, $\sigma_0=1.0$. In the right-hand panel all runs with nonuniform resistivity $\eta = 10^{-4}(1+\Delta_{ei}^2 J)$ are shown, with $\Delta_{ei}^2 = 0$ for run Fs, $\Delta_{ei}^2 = 0.001$ for run Es, $\Delta_{ei}^2 = 0.01$ for run Ds, $\Delta_{ei}^2 = 0.1$ for run Cs and $\Delta_{ei}^2 = 1$ for run Bs.}
\label{fig:gammavst}
\end{center}
\end{figure*}

Unperturbed runs G (solid yellow line) and Gs (dashed yellow line) show only an initial acceleration phase that is due to a resistive electric field that forms in the flux ropes. Current sheets do not form in these runs and therefore the acceleration is slower, and $\Gamma_{max}$ is smaller than for the perturbed runs. Here, the effect of destroying particles that leave the current sheet, and injecting thermal particles, is visible in the maximum Lorentz factor that is reached for run Gs, that is approximately four times smaller than for run G at the final time. This difference quantifies the effect of the bulk heating of particles in the flux tubes. To confirm that this is a resistive effect, a simulation is conducted in the ideal SRMHD module of {\tt BHAC}, such that $\eta = 0$ by definition and $\mathbf{E} = -\mathbf{u} \times \mathbf{B}/\Gamma$, with $\mathbf{u}$ and $\Gamma$ the fluid momentum and Lorentz factor. In the ideal case Gsi, the Lorentz factor is bounded to $\Gamma_{max} \lesssim 1.7$ on the time scales considered, since no resistive electric field forms. There is no exponential growth phase and particles only mildly accelerate in the unmoving flux ropes due to the ideal electric field $\mathbf{E} = -\mathbf{u} \times \mathbf{B}/\Gamma$. This confirms that even in unperturbed runs, the resistive electric field is the main acceleration mechanism.

In the middle panel of Figure \ref{fig:gammavst}, the effects of $\sigma_0$ and plasma-$\beta_0$ are quantified for runs with uniform resistivity $\eta = 5\times10^{-5}$. The larger magnetization of run Bs ($\beta_0=0.1$, $\sigma_0=3.3$), compared to run Js ($\beta_0=0.5$, $\sigma_0=0.9$) results in a slightly faster initial acceleration phase. Increasing the magnetization from $\sigma_0=0.9$ for run Js, to $\sigma_0=1.0$ for run Ks, results in a minor increase in the slope of $\Gamma_{max}$, however still yields a less fast acceleration than for run Bs with lower plasma-$\beta$. Runs Js and Ks suggest that a smaller $\sigma$ limits maximum energy of the particles, in accordance with the PiC findings of \cite{SironiPorth}.

In the right-hand panel of Figure \ref{fig:gammavst}, the evolution of $\Gamma_{max}$ for all nonuniform resistivity runs are shown, and compared to run Fs with uniform resistivity $\eta = 10^{-4}$. For run Bs, with $\Delta_{ei}^2=1$, the maximum Lorentz factor $\Gamma_{max}$ grows fast, but then reaches an approximately constant maximum after $t_c \approx 1$. This corresponds to the point where the current sheet breaks-up and rapidly diffuses due to the strongly enhanced resistivity. In run Cs ($\Delta_{ei}^2=0.1$), the largest $\Gamma_{max}$ is reached at $t_c \approx 10$, after which the acceleration decays due to the diffusion of the current sheet. In runs Ds ($\Delta_{ei}^2=0.01$) and Es ($\Delta_{ei}^2=0.001$), the resistivity is only mildly enhanced, and the maximum Lorentz factor follows the evolution of run Fs with uniform resistivity. The maximum Lorentz factor scales with the maximum of the nonuniform resistivity here, and only in the far nonlinear phase ($t_c \gtrsim 18$) a decrease of $\Gamma_{max}$ is observable for run Ds. This confirms again that the resistive electric field $\mathbf{E} \sim \eta \mathbf{J}$ is the main acceleration mechanism.

\section{Conclusions}
We evolve charged test particles in electromagnetic fields obtained from SRRMHD simulations of 2D merging Lundquist tubes for a range of Lundquist numbers, plasma-$\beta$ and magnetizations. We find high-energy power-law tails forming due to acceleration in the current sheet in cases with $\eta \leq 10^{-4}$. The maximum Lorentz factor of the particle ensemble is proportional to the resistive electric field (and hence proportional to the resistivity). After the initial acceleration in the forming current sheet, a secondary acceleration regime is reached with a larger variability of the Lorentz factor.

We find that electrons accelerate to medium Lorentz factors in the flux tubes and to high Lorentz factors in the forming current sheet in between the flux tubes. The latter is in agreement with PiC results of \cite{SironiPorth}, but the bulk heating in the original flux tubes is not. By conducting an ideal SRMHD simulation (i.e. $\eta=0$) we have shown that the bulk acceleration in the current sheet is mainly due to the too large resistivity in the ambient plasma and in the two flux tubes.

With a realistic spatiotemporally dependent resistivity and a very low base resistivity, microscopic effects like plasmoid formation and particle acceleration can be restricted to small parts of the domain where a large current density is observed. 
By prescribing a very small background resistivity $\eta \ll 10^{-5}$ and a current-dependent resistivity below the plasmoid threshold, $\eta < 5 \times 10^{-5}$ in future applications, we expect plasmoid formation and particle acceleration in reconnection regions, and a thermal particle distribution in the ambient and in the flux tubes. 

In this work particles are evolved using the special relativistic Boris scheme, neglecting curved spacetime. In accreting black hole systems, the ejection of plasmoids and the associated flaring might occur close to the black hole event horizon. The special relativistic particle integrator used here is incapable of taking the effect of gravity into account. In combination with the capabilities documented in \cite{bacchini2018iau} and \cite{bacchini2018} for particle integrators in curved spacetime, we prepare for future applications of accretion physics and relativistic jet dynamics, and the reconnection physics and particle acceleration associated with them.

\section*{Acknowledgements}
This research was supported by project GOA/2015-014 (2014-2018 KU Leuven). OP is supported by the ERC synergy grant `BlackHoleCam: Imaging the Event Horizon of Black Holes' (Grant No. 610058). 
The computational resources and services used in this work were provided by the VSC (Flemish Supercomputer Center), funded by the Research Foundation Flanders (FWO) and the Flemish Government - department EWI.
BR would like to thank Lorenzo Sironi for useful discussions.

\end{document}